\documentclass[a4paper,11pt]{article}
\pdfoutput=1 

\usepackage{jcappub} 

\usepackage[T1]{fontenc} 
\usepackage{mathrsfs}
\usepackage{amsmath,amssymb}
\usepackage{slashed} 
\usepackage{bbm} 
\usepackage{accents} 
\usepackage{soul}

\title{Dynamical Screening Effects on \\ Big Bang Nucleosynthesis}

\author[a]{Eunseok Hwang}
\author[b,1]{Dukjae Jang \note{Corresponding author.}}
\author[a]{Kiwan Park} 
\author[c,d]{\\Motohiko Kusakabe}
\author[c,d,e]{Toshitaka Kajino}
\author[f,d]{\\A. Baha Balantekin}
\author[g,d]{Tomoyuki Maruyama}
\author[b,h]{Chang-Mo Ryu}
\author[a,c,d]{Myung-Ki Cheoun}

\affiliation[a]{Department of Physics and OMEG Institute, Soongsil University, Seoul 156-743, Republic of Korea}
\affiliation[b]{Center for Relativistic Laser Science, Institute for Basic Science (IBS), Gwangju 61005, Republic of Korea}
\affiliation[c]{School of Physics and International Research Center for Big-Bang Cosmology and Element Genesis, Beihang University, Beijing 100083, China}
\affiliation[d]{National Astronomical Observatory of Japan, 2-21-1 Osawa, Mitaka, Tokyo 181-8588, Japan}
\affiliation[e]{The University of Tokyo, Bunkyo-ku, Tokyo 113-0033, Japan}
\affiliation[f]{Physics Department, University of Wisconsin-Madison, 1150 University Avenue, Madison, Wisconsin 53706, USA}
\affiliation[g]{College of Bioresource Sciences, Nihon University, Fujisawa 252-0880, Kanagawa-ken, Japan}
\affiliation[h]{Department of Physics, Pohang University of Science and Technology, Pohang 790-784, Republic of Korea}

\emailAdd{hwangeunseok94@gmail.com}
\emailAdd{djjang2@ibs.re.kr}
\emailAdd{pkiwan@gmail.com}
\emailAdd{kusakabe@buaa.edu.cn}
\emailAdd{kajino@buaa.edu.cn}
\emailAdd{baha@physics.wisc.edu}
\emailAdd{maruyama.tomoyuki@nihon-u.ac.jp}
\emailAdd{ryu201@postech.ac.kr}
\emailAdd{cheoun@ssu.ac.kr}

\abstract{A moving ion in plasma creates a deformed electric potential depending on the ion velocity, which leads to the distinct screening effect compared to the standard static Salpeter formula. In this paper, adopting the test charge method, we explore the dynamical screening effects on big bang nucleosynthesis (BBN). We find that the high temperature in the early universe causes the ion velocity to be faster than the solar condition so that the electric potential is effectively polarized. However, the low density of background plasma components significantly suppresses the dynamical screening effects on thermonuclear reaction rates during the BBN epoch. We compare our results with several thermonuclear reaction rates for solar fusion considering the dynamical screening effects. Also, we discuss the additional plasma properties in other astrophysical sites for the possible expansion from the present calculation in the future.}

\begin{document}
\maketitle
\flushbottom

\section{Introduction}
\label{interoduction}
One of fundamental characteristics in plasma is the Debye shielding. The electron cloud surrounding the charge of an ion screens other nuclear charges far from the own radius, approximately given as Debye radius $\lambda_D$. In a nuclear reaction, the screening effect reduces the Coulomb barrier, so that the penetration probability is enhanced. Under the weak screening condition, the Salpeter formula \cite{1954AuJPh...7..373S} well describes the static screening effects for the thermonuclear reactions, which result in the enhancement factor $f_{en}$ in terms of given temperature $T$ as follows:
\begin{eqnarray}
f_{en} = \exp \left[ \frac{Z_1Z_2 e^2}{\lambda_{D,e} T} \right],
\label{salpeter}
\end{eqnarray}
where $Z_i$ denotes the charge number of species $i$, $e$ is the electronic charge, and the Debye radius $\lambda_{D,e}$ is defined as
\begin{eqnarray}
\lambda_{D,e} = \sqrt{\frac{T}{4\pi n_e e^2}},
\end{eqnarray}
depending on the temperature and the electron number density $n_e$.

Since the screening effects change the thermonuclear reaction rates, the Salpeter form has been exploited widely in nucleosynthesis such as the stellar nuclear fusion \cite{2013CoPP...53..397P},  presupernova (preSN) \cite{2007A&A...463..261L},  and core-collapsing supernova \cite{2009MNRAS.400..815L}. Among various studies, in particular, big bang nucleosynthesis (BBN) has been a good testbed for the screening effects owing to a relatively small number of main reactions and precise observational data based on cosmic microwave background (CMB) study and astronomical spectroscopy. The first application of the Salpeter formula to the BBN shows that the screening effects hardly affect the primordial abundances because the density in the BBN environment is too low to reveal the effective charge shielding \cite{2011PhRvC..83a8801W}.  However, it was pointed out that the relativistic corrections are considerable in the BBN environment due to enough high temperatures ($10^8\,{\rm K} \lesssim T \lesssim 10^{10}\,{\rm K}$) to maintain the electron-positron plasma. Then, replacing the Debye radius with the Thomas-Fermi one, the relativistic correction of screening effects on BBN was performed, but the result still shows insignificant fractional changes of primordial abundances within the order of $10^{-3}$ \cite{2016PhRvC..93d5804F}. Furthermore, a study of the relativistic screening effects with the primordial magnetic field (PMF) shows remarkable changes of primordial abundances. In reference \cite{2020PhRvD.101h3010L}, the PMF was consistently treated in the cosmic expansion rate, the temperature evolution, and screening correction for weak reactions, which are constrained with observational data of the primordial abundances.

Another question on the screening remains in dynamical effects by a moving ion. When a test charge moves with an intermediate velocity that is enough to react with background charges in plasma, the Coulomb potential forms a deformed shape \cite{1981JPlPh..25..225W,1993JETP...77..910T}, which leads to an aspherical Debye shielding. This dynamical effect on the electron screening was firstly mentioned in reference \cite{1977ApJ...212..513M} and studied for the thermonuclear reaction rates by C. Carraro et al. \cite{1988ApJ...331..565C} in the context of the solar neutrino problems \cite{2016RvMP...88c0502M, 2016RvMP...88c0501K}. For this argument, reference \cite{1998ApJ...496..503G} suggested that the dynamical screening effects vanish arguing that the factorization between kinetic and interaction parts in the Gibbs probability guarantees the independence of interaction term from the kinetic energy (See also references \cite{2002A&A...383..291B} and \cite{1997RvMP...69..411B}.). On the other hand, reference \cite{2000ApJ...535..473O} disputed the factorizability pointing out that the dynamical potential involves the particle velocity. Since the electric potential in plasma depends on the velocity of a moving ion, the dynamical effects on screening should be present when charges move with large enough velocities. However, whether they are realized in astrophysical conditions or not is currently an open question. Hence, for the dynamical screening effects, subsequent studies in solar fusion have investigated various methods such as molecular dynamic simulation \cite{1999PhR...311...99S, 2010Ap&SS.328..153M, 2011ApJ...729...96M}, kinetic equation \cite{2000A&A...356L..57T}, and field theoretical approaches \cite{2017PhRvD..95k6002Y}.

In this study, we investigate the dynamical screening effects on BBN. Compared to the solar condition, the BBN occurs in a state of low density, but high temperature. Also, the rapid cooldown of the early universe changes the number density of plasma components so that the BBN plasma undergoes the transition from an electron-positron-ion (EPI) to electron-ion (EI) plasma. Because of the evolving environment, the screening potential for the thermonuclear reaction rates is given in terms of the cosmic time. This paper shows a time-evolving dielectric function for the longitudinal mode and effects of the dynamical screening on the nuclear reaction rates in BBN adopting the test charge method used in reference \cite{1988ApJ...331..565C}.

The rest of this paper is organized as follows. In Section 2, we begin by deriving the dielectric permittivity for a moving test charge in Coulomb potential and apply it to the BBN condition. Based on the obtained Coulomb potential, Section 3 discusses the thermonuclear reaction rates with dynamical screening effects. Showing the enhancement factor for main BBN reactions, we also investigate those effects on primordial abundances. In Section 4, we make a conclusion of this study, discussing results and a future implication for other astrophysical environments.

\section{Coulomb potential for moving ions in the BBN epoch}
From the Poisson equation for a moving test charge with time $t$, the Coulomb potential can be written as \cite{1988ApJ...331..565C}
\begin{eqnarray}
\phi ({\bf r}-{\bf v}t) = \frac{Z_0 e}{2\pi^2} \int \frac{d {\bf k}}{k^2} \frac{e^{i {\bf k} \cdot ({\bf r}-{\bf v}t)}}{\epsilon_l ({\bf k}, {\bf k \cdot v})},
\label{eq_pot}
\end{eqnarray}
where $Z_0$ denotes charge number of the test charge, ${\bf r}$ the spherical coordinate in the rest frame, and ${\bf v}$ the velocity of the moving test charge. The longitudinal mode of dielectric permittivity $\epsilon_l({\bf k}, \omega)$, as a function of wavevector ${\bf k}$ and frequency $\omega$, is derived by the first order perturbation of the Vlasov-Maxwell equation with unmagnetized equilibrium plasma \cite{1981phki.book.....L}. When we consider the electron ($e^-$), positron ($e^+$), $^1$H, and $^4$He as components in the early universe plasma, $\epsilon_l({\bf k}, \omega)$ for test charge moving along $x$-direction is given as
\begin{eqnarray}
\epsilon_l({\bf k}, \omega) = 1 &-& \frac{4\pi e^2 g_e}{k} \int_{-\infty}^{\infty} \frac{d f_e (p_x)}{d p_x} \frac{dp_x}{kv_x - \omega} +\frac{1}{(k\lambda_{D, ^{1}{\rm H}})^2}  \left[ 1 + F \left( \frac{\omega}{\sqrt{2} k v_{T, ^{1}{\rm H}}} \right)  \right]  \nonumber \\[12pt]
&+& \frac{1}{(k\lambda_{D, ^4{\rm He}})^2} \left[ 1 + F \left( \frac{\omega}{\sqrt{2} k v_{T, ^4{\rm He}}} \right)  \right].
\label{eq_perm}
\end{eqnarray}
The first and second terms, respectively, stand for the dielectric permittivity in vacuum and the net current of $e^-$ and $e^+$. For the statistical degrees of freedom for $e^-$ and $e^+$, we take $g_e=2$ and $f_e(p_x)$ denotes the integration of the momentum distribution for the net electron over $y$ and $z$ components, i.e., $f_e(p_x) \equiv \int f(p) dp_y dp_z$. We assume that all species have same temperature under the thermal equilibrium condition and adopt Fermi-Dirac distributions for the $e^-$ and $e^+$, which defines $f_e$ as follows:
\begin{eqnarray}
f_e &\equiv& f_{e^-} - f_{e^+} = \left[ \frac{1}{\exp \left(  \frac{\sqrt{{p}^2 + m_{e}^2} - \mu}{T} \right)  + 1 } - \frac{1}{\exp \left( { \frac{\sqrt{{p}^2 + m_{e}^2} + \mu}{T}} \right)  + 1}  \right] ,
\label{eq_net_numb}
\end{eqnarray}
where $p$ is the momentum, $m_{e}$ the mass of $e^-$ (or $e^+$), and $\mu$ the chemical potential determined by the charge neutrality condition of $n_{e^-} = n_{e^+} + n_{\rm H} + 2 n_{^4{\rm He}}$. In time-evolving BBN condition, the chemical potential $\mu$ is varied with cosmic time \cite{2018PhR...754....1P}. 

In the dielectric function (equation (\ref{eq_perm})), the third and fourth terms result from the contributions by the $^1$H and $^4$He. Here we neglect contributions of other nuclei such as D, $^3$He, $^7$Li, and $^7$Be due to their small numbers. For each species, we use the Debye radius and thermal velocity as $\lambda_{D,i} = \sqrt{T/(4\pi n_i Z_i^2 e^2)}$ and ${v_{T,i}}=(T/m_i)^{1/2}$, respectively. Nuclear mass fraction $X_i$ is related to its number density $n_{i} = X_i \eta n_\gamma/ A_i $, where $\eta$, $n_\gamma$, and $A_i$ stand for the baryon-to-photon ratio, number density of photon, and mass number of species $i$, respectively. To obtain the mass fraction of protons and $^4$He, we use the updated BBN calculation code in references \cite{1992STIN...9225163K,  1993ApJS...85..219S} with reactions in references \cite{2004ADNDT..88..203D, 2016ApJ...831..107I}, and adopt parameters as follows: neutron lifetime $\tau_n = 879.4\,s$ \cite{2018PhRvD..98c0001T}; $\eta=6.037 \times 10^{-10}$ \cite{2016A&A...594A..13P}; effective neutrino number $N_{\rm eff}=3.046$ \cite{2005NuPhB.729..221M}. Function $F(x)$ in equation \eqref{eq_perm} is defined as \cite{1981phki.book.....L}
\begin{eqnarray}
F(x) = \frac{x}{\pi^{1/2}} \int_{-\infty}^{\infty} \frac{e^{-z^2}}{z-x}dz + i \pi^{1/2} x e^{-x^2}.
\label{eq_fx}
\end{eqnarray}
Note that $\epsilon_l({\bf k}, \omega)$ goes to unity corresponding to free space when all components are absent. In other words, the contribution of plasma components makes the permittivity deviate from the value in free space, which affects the shape of Coulomb potential.

\begin{figure}[t]
\begin{center}
\includegraphics[width=11 cm]{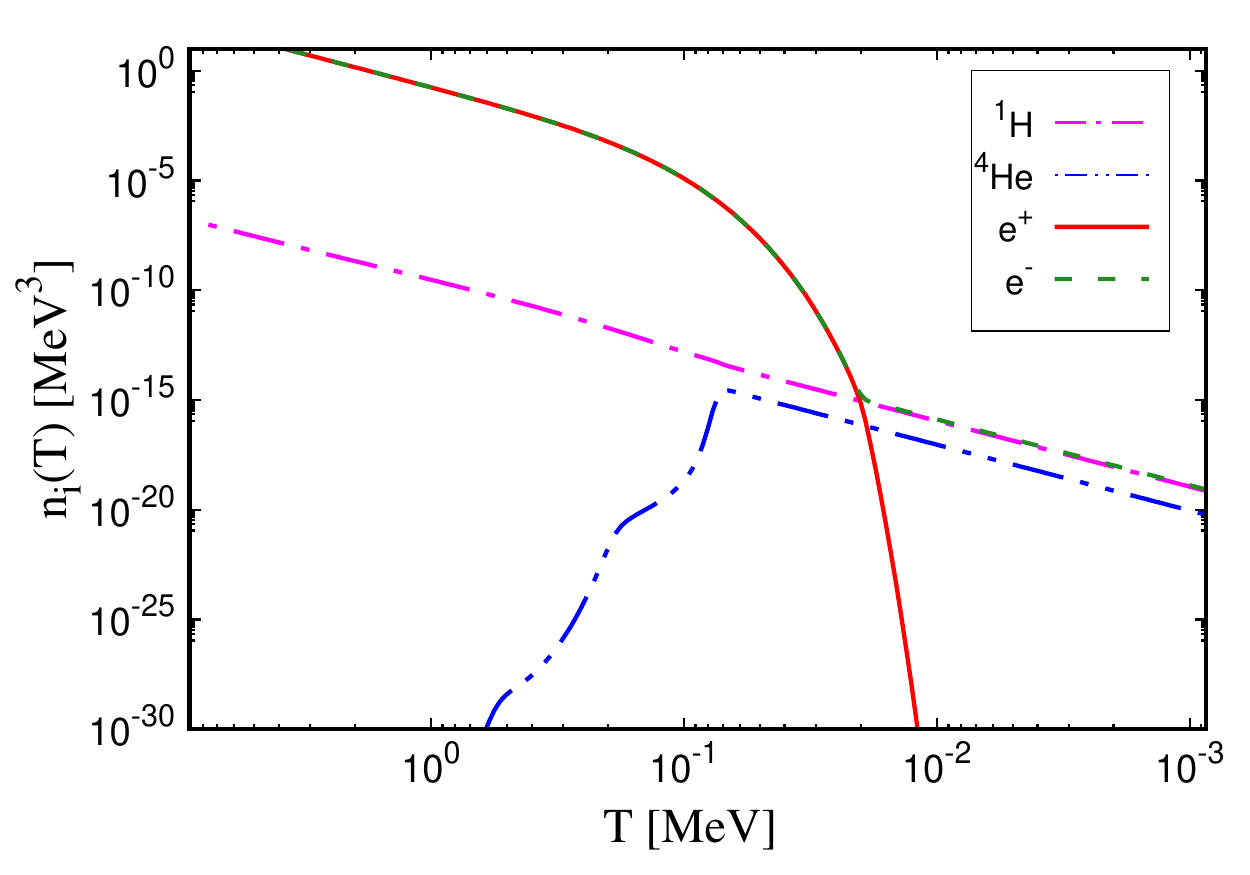}
\end{center}
\caption{The number densities of plasma components as a function of BBN temperature. Magenta-dashed-single dotted, blue-dashed-double dotted, red-solid, and green-dashed lines denote the number densities of $^1$H, $^4$He, $e^+$, and $e^-$, respectively.}
\label{fig1}
\end{figure}

Figure \ref{fig1} shows the number densities of plasma components during the BBN epoch. In the high temperature region corresponding to early cosmic time, the number density of relativistic $e^-$ and $e^+$ is proportional to $T^3$, and they are dominant components. Over the cosmic time, the universe rapidly cools down by the adiabatic expansion, and $n_{e^-}$ and $n_{e^+}$ drastically decrease when $e^-$s and $e^+$s become non-relativistic at $T \lesssim m_e$. On the other hand, the $^4$He abundance increases by the nucleosynthesis as follows. In the early phase, BBN condition maintains enough high temperature to allow the nuclear statistical equilibrium (NSE). The NSE continues to increase $^4$He up to $T=0.6\,{\rm MeV}$, where the relatively slow increase of $^3$H and $^3$He slows down the reaction rate of $^3{\rm H}(p,\gamma)^4{\rm He}$ and $^3{\rm He}(n,\gamma)^4{\rm He}$. By the slow production rate, $^4$He is decoupled from NSE and follows the NSE curve of $^3$H and $^3$He. Similarly, at $T=0.2\,{\rm MeV}$, the tardy growth of deuterium slows down reaction rates of $^2{\rm H}(n,\gamma)^3{\rm H}$ and $^2{\rm H}(p,\gamma)^3{\rm He}$, which lead to the decoupling of $^3$H and $^3$He from the NSE. At $T=0.07\,{\rm MeV}$, increase of $^3$H, $^3$He and $^4$He following the NSE curve of deuterium stops by decoupling of deuterium from NSE. Consequently, those three kinds of decoupling from NSE at $T=0.6\,{\rm MeV}$, $0.2\,{\rm MeV}$, and $0.07\,{\rm MeV}$ make the broken lines for $^4$He shown in figure \ref{fig1} (See also references \cite{1993ApJS...85..219S, 1996RPPh...59.1493S} for details.). Due to this synthesis of $^4$He, a difference between $n_{\rm H}$ and $n_{e^-}$ remains so as to satisfy the charge neutrality condition. The change of dominant plasma components transforms the early universe from EPI to EI plasma, and as a result, the dielectric permittivity becomes cosmic time dependent.

\begin{figure*}[t]
\centering
\includegraphics[width=7.5cm]{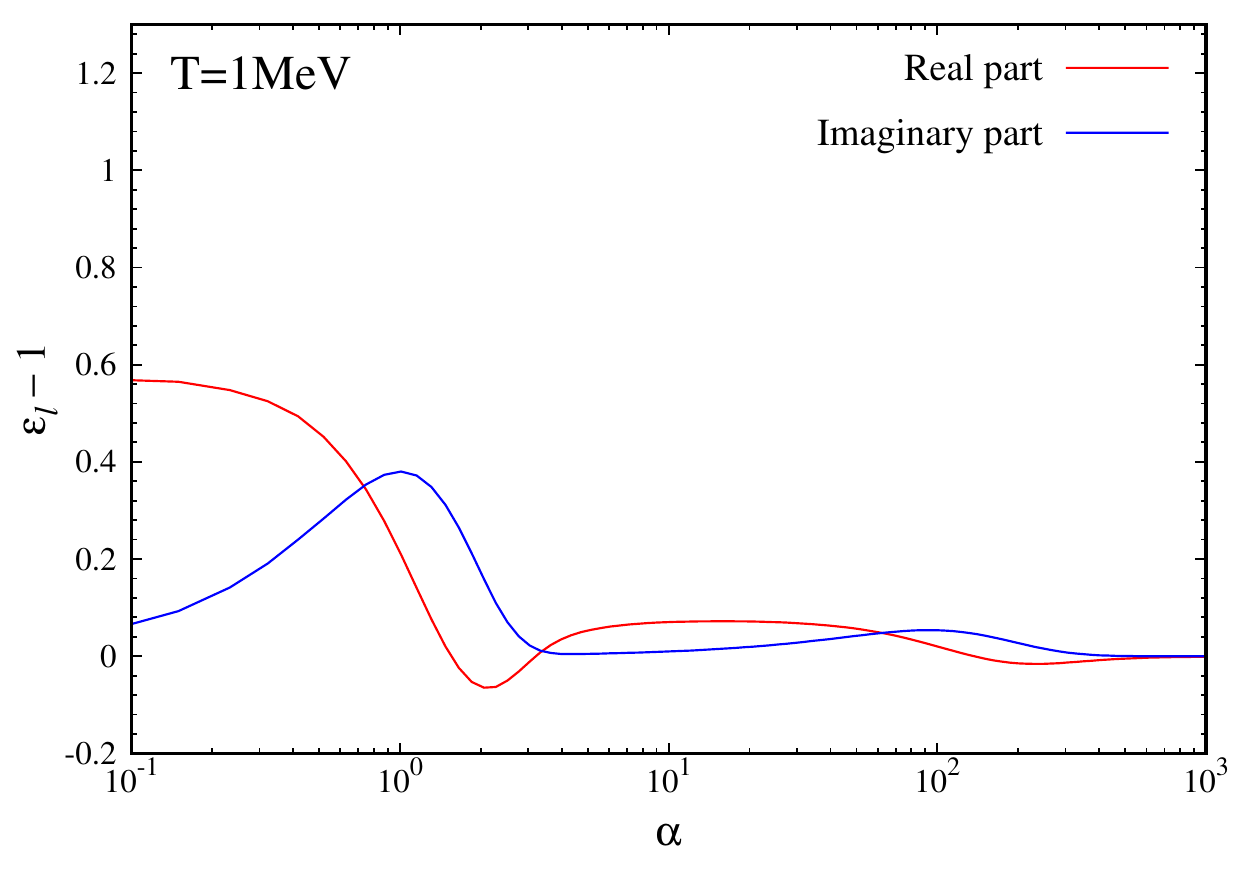}
\includegraphics[width=7.5cm]{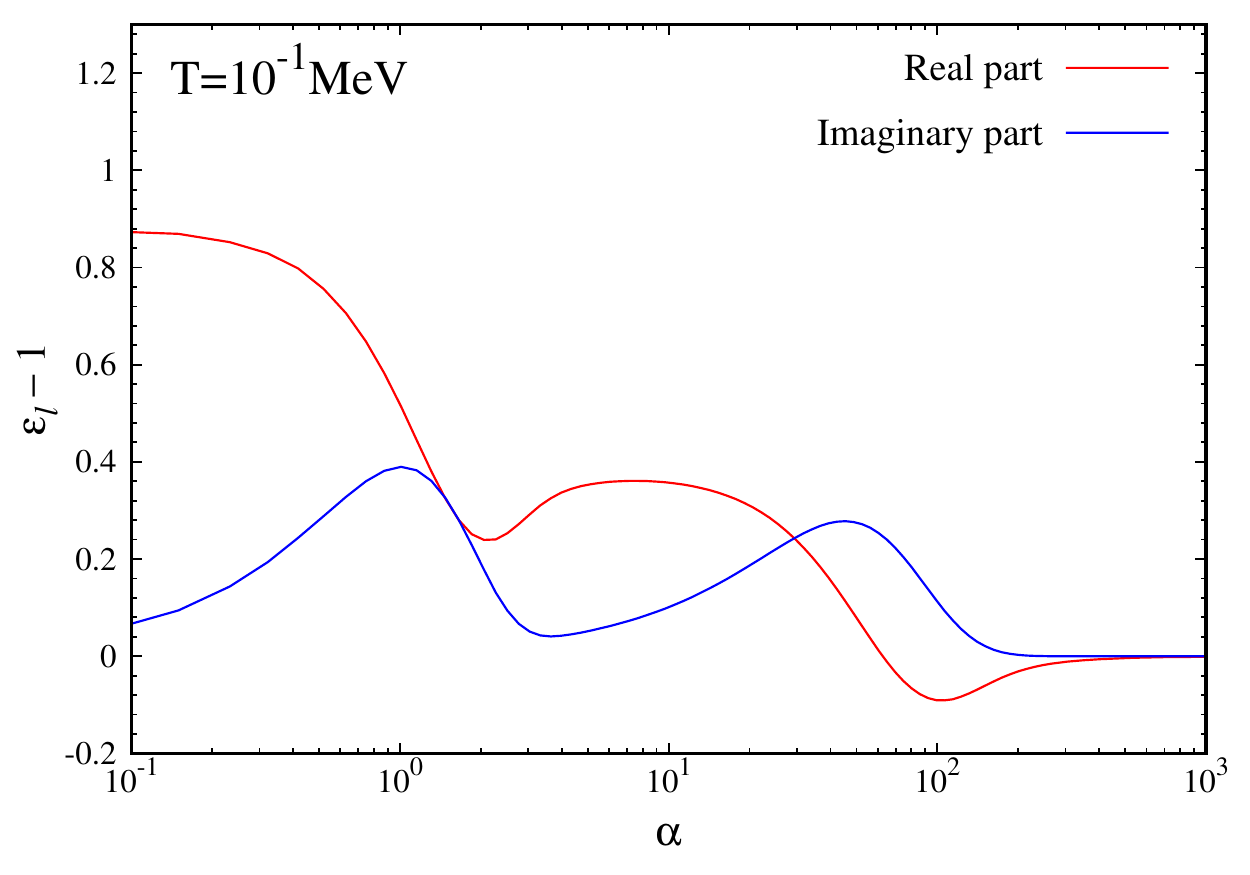}
\includegraphics[width=7.5cm]{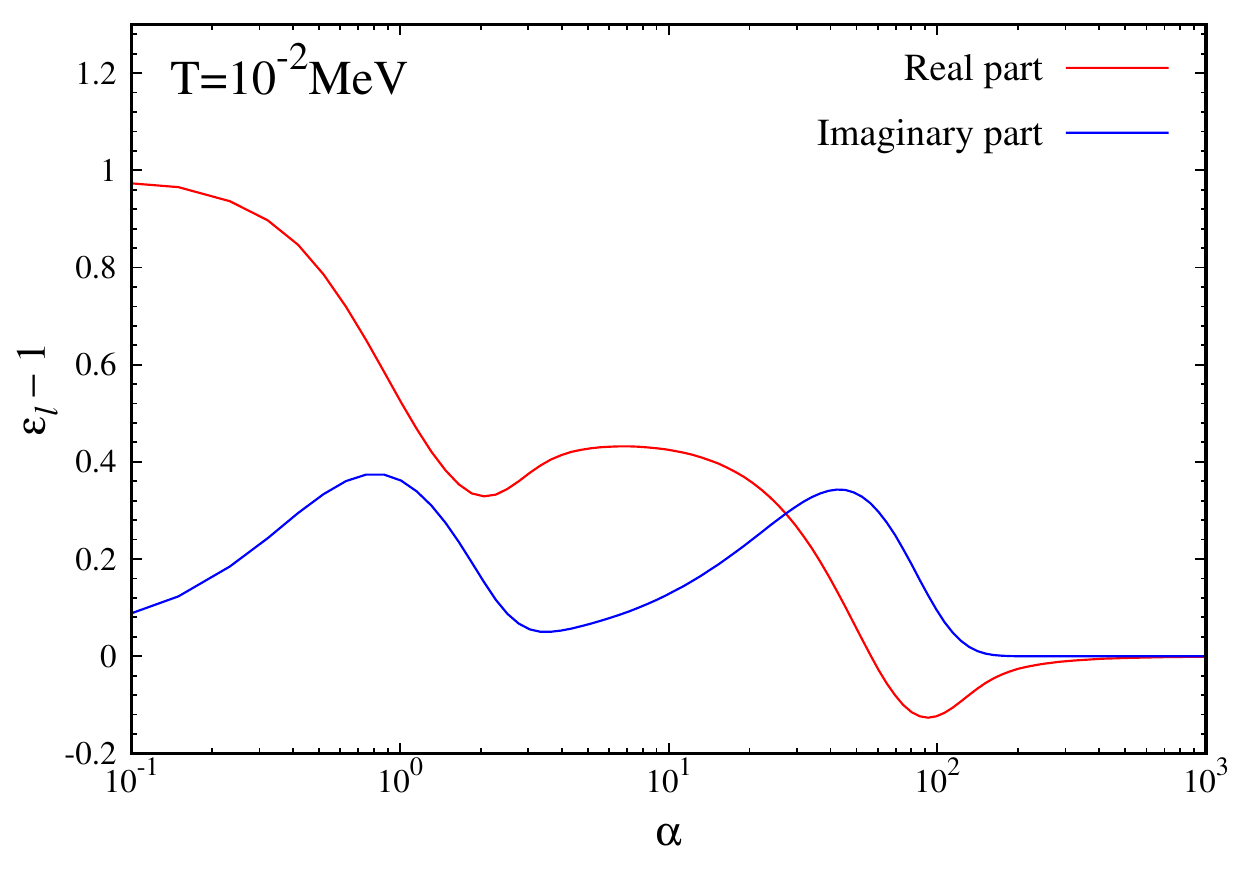}
\includegraphics[width=7.5cm]{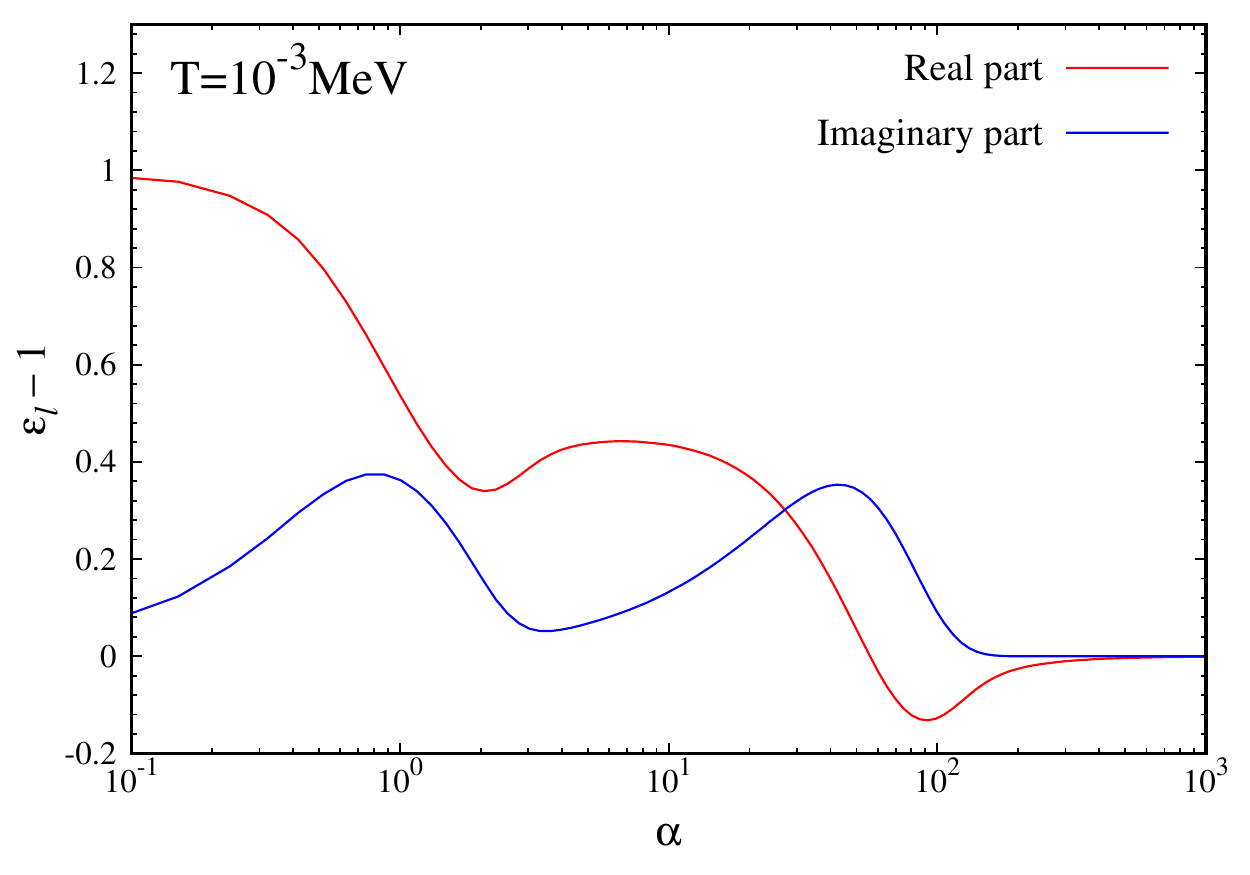}
\caption{Difference of dielectric permittivity between BBN plasma and free space i.e., $\epsilon_l-1$, as a function of $\alpha\ (=v/v_{T,^1{\rm H}})$ at $T=1$\,MeV (upper-left), $T=0.1$\,MeV (upper-right), $T=0.01$\,MeV (lower-left), and $T=0.001$\,MeV (lower-right). Red and blue solid lines denote the real and imaginary parts of the dielectric permittivity, respectively.}
\label{fig2}
\end{figure*}

Figure \ref{fig2} shows the deviation of $\epsilon_l({\bf k}, \omega)$ from the unity as a function of $\alpha \equiv v/v_{T, ^1\rm H}$. When the proton has high velocity ($\alpha \ge 10^2$), as shown in all panels in figure \ref{fig2}, the dielectric permittivity converges to unity, i.e., one in free space. It means that the ion velocity is so fast that background plasma cannot react to the moving ion. In this very high velocity region, we should consider the relativistic correction of the moving ion, but the typical temperatures of astrophysical environments hardly allow the high velocity of ions due to large mass. On the other hand, near the thermal velocity region, the dielectric permittivity relies on the background property depending on temperature.

At $T=1$\,MeV, relativistic $e^-$s and $e^+$s dominate the BBN plasma as explained above. At this moment, the near-equality of number densities $e^-$s and $e^+$s leads the net current to become null, so the permittivity is rarely changed. This dominance of $e^-$ and $e^+$ continues to $T=0.1$\,MeV, and the change of permittivity is not remarkable at $T=0.1$\,MeV (See upper panels in figure \ref{fig2}.). On the other hand, by decreasing of temperature at $T \sim 0.01$\,MeV, the number densities of $e^-$s and $e^+$s rapidly reduce to a level comparable to that of baryons. As neutrons decay during the BBN, electrons are produced and $e^+$ number density reduces in order to satisfy the charge neutrality. When $n_{e^-}$ decreases to $\sim n_{\rm H}$, the difference between $n_{e^-}$ and $n_{e^+}$ becomes significant. While positrons continue to annihilate, the electron annihilation freezes out. Then, a deviation of the permittivity from unity develops. However, freeze-out of nuclear reactions does not change the nuclear abundance after $T=0.01$\,MeV, which makes the deviation at $T=0.001$\,MeV similar to the $T=0.01$\,MeV case. Therefore, the lower two panels in figure \ref{fig2} show the similar deviations of the permittivity from that of the free space.

In the imaginary part related to a damping of electric potential, an oscillation behavior stems from the function $F(x)$ defined in equation \eqref{eq_fx}. At $T=1\,{\rm MeV}$, the $^4$He synthesis does not start yet, and the fourth term in equation \eqref{eq_perm} can be omitted. Then, the only one peak by the proton ion is shown in the upper-left panel of figure \ref{fig2}. After the $^4$He synthesis, the contribution of $^4$He ion to the imaginary part appears, by which the second peak is seen in other panels.

\begin{figure*}[t]
\centering
\includegraphics[width=13 cm]{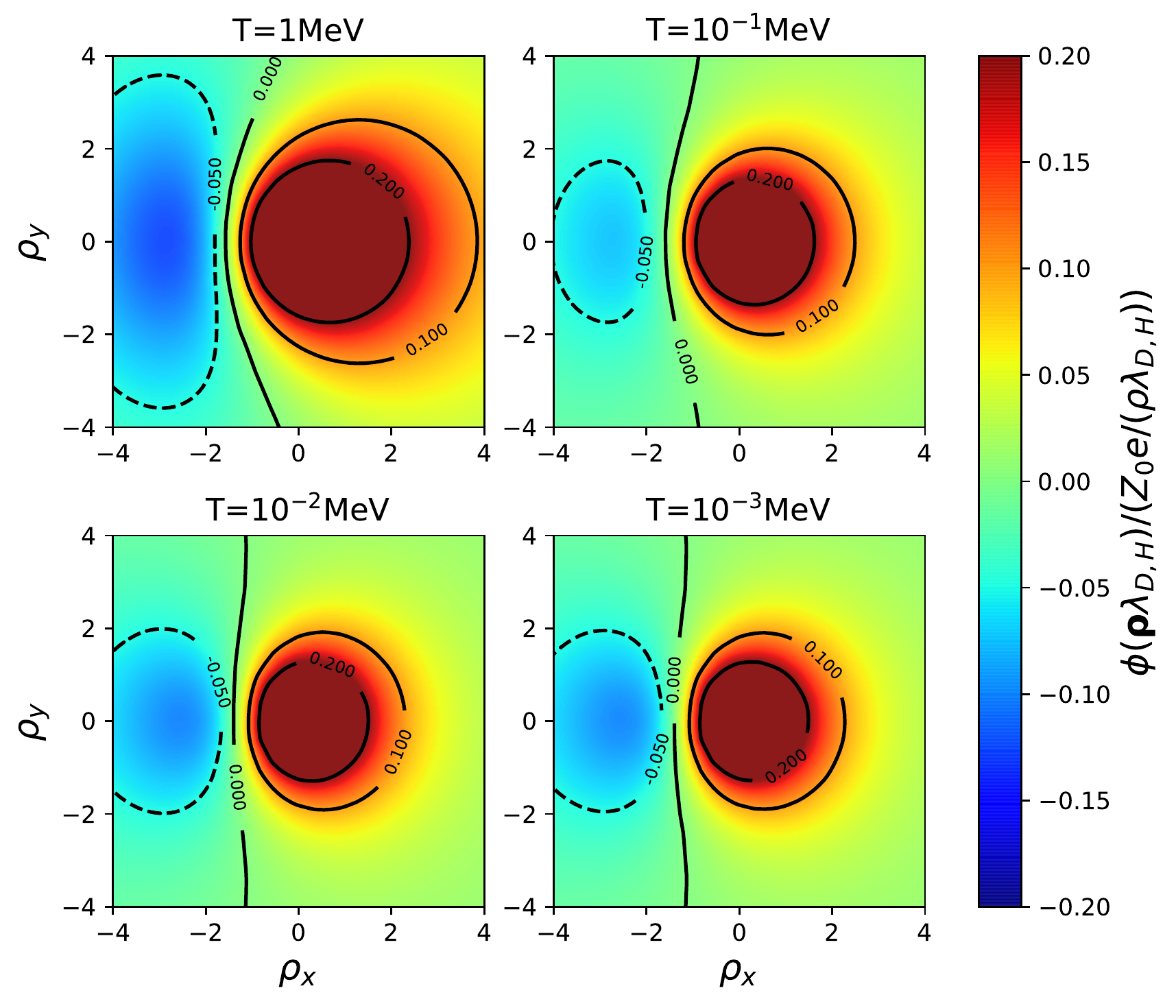}
\caption{Normalized electric potential for a moving ion under the BBN condition, i.e., $\phi({\boldsymbol{\rho}} \lambda_{D,H})/(Z_0 e/(\rho \lambda_{D,H}))$ where ${\boldsymbol{\rho}} \equiv {\bf R}/\lambda_D$. As a test charge, we adopt a proton moving along $x$-direction in the dimensionless real space of $\rho_x\ (\equiv x/\lambda_{D,H})$ and $\rho_y\ (\equiv y/\lambda_{D,H})$ with the thermal velocity, i.e., $v=\sqrt{T/m_{^1{\rm H}}}$.}
\label{fig3}
\end{figure*}

Adopting the obtained dielectric permittivity, we calculate the Coulomb potential for a moving proton. For convenience, we perform a transformation from the fluid rest frame to the moving proton frame, which leads equation \eqref{eq_pot} to
\begin{eqnarray}
\phi ({\bf R}) = \frac{Z_0 e}{2\pi^2} \int \frac{d {\bf k}}{k^2} \frac{e^{ i {\bf k \cdot {\bf R}}}}{\epsilon_l ({\bf k}, {\bf k \cdot v})},
\label{pot_trans}
\end{eqnarray}
where ${\bf R} \equiv {\bf r}-{\bf v}t$. Figure \ref{fig3} shows the electric potential normalized by Coulomb potential, i.e., $\phi({\bf R})/(Z_0 e/R)$, for a proton moving along the $x$-axis in the plane of the two dimensional coordinate as $({\rho}_x \equiv x/\lambda_{D,H}, \rho_y \equiv y/\lambda_{D,H})$. According to equation \eqref{pot_trans}, the Coulomb potential for a moving test charge depends on the dielectric permittivity as well as ion thermal velocity or temperature. Here the velocity of the proton is set to the thermal velocity. The dependence of the potential on the velocity and the dielectric permittivity implies that the shape of the potential also evolves with cosmic temperature.

In the non-relativistic regime, it is known that the faster the test charge is, the more deformed the shape of the electric field is from a spherical shape \cite{1981JPlPh..25..225W, 1988ApJ...331..565C}. At $T=1\,{\rm MeV}$, the thermal velocity of the proton is the highest among the panels in figure \ref{fig3}, which effectively polarizes the background charge. As a result, at $T=1\,{\rm MeV}$, the valley in the backward direction of the moving charge has the minimum value and the shape of the potential is largely deformed, despite the small deviation of dielectric permittivity. On the other hand, the potential in the forward direction is increased. At $T=0.1\,{\rm  MeV}$, the dielectric permittivity shows a still similar pattern although the thermal velocity is smaller. The reduced velocity makes the distortion of the potential relatively small.

\begin{figure*}[t]
\centering
\includegraphics[width=11 cm]{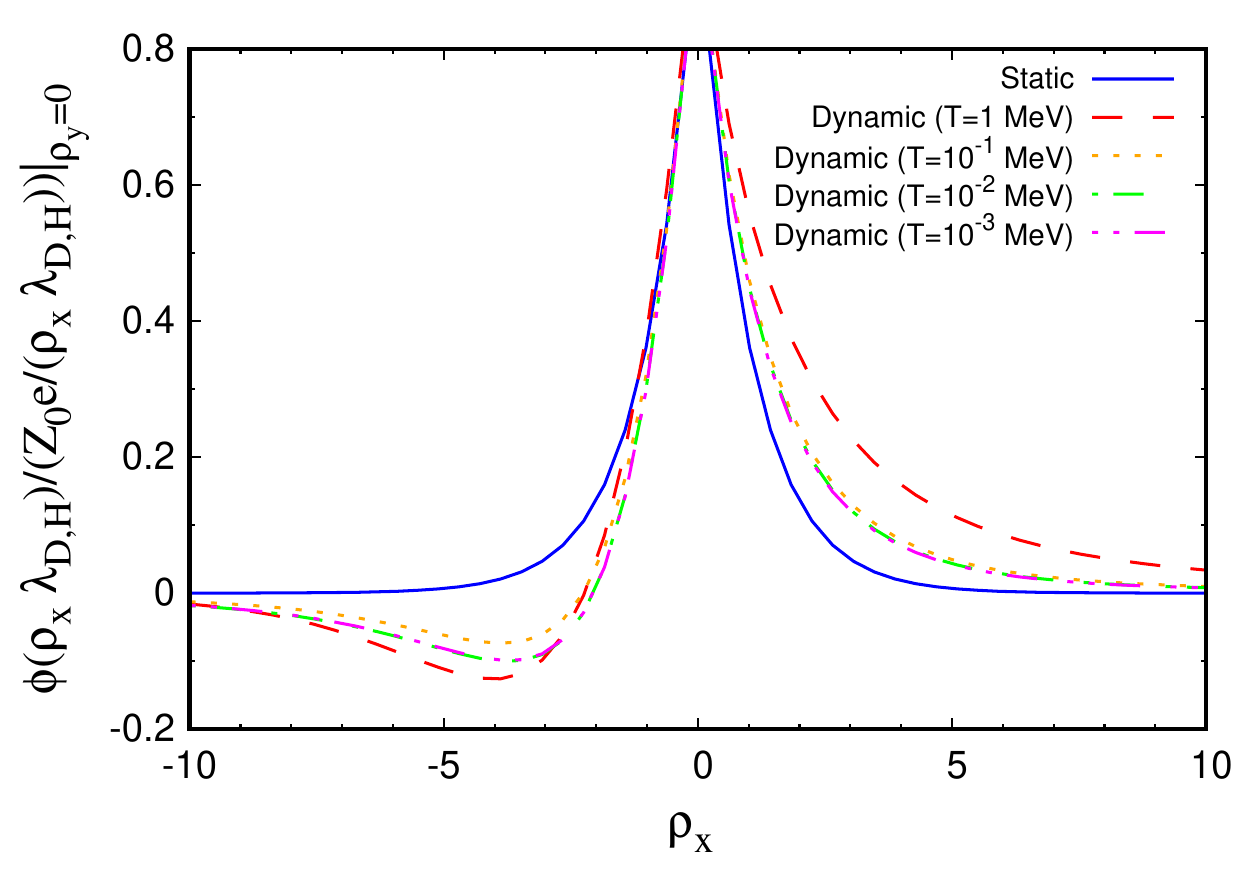}
\caption{Normalized electric potential for $\rho_y=0$. Blue-solid line shows the electric potential for a static charge. Red-dashed, orange-dotted, green-dashed-single dotted, and magenta-dashed-double dotted lines denote the potential for dynamical charges at $T=1\,{\rm MeV}, 0.1\,{\rm MeV}, 0.01\,{\rm MeV}$, and $0.001\,{\rm MeV}$, respectively.}
\label{fig3_add}
\end{figure*}

At $T=0.01\,{\rm MeV}$ (the left-bottom panel of figure \ref{fig3}), the deviation of dielectric permittivity is larger by growing up of the heavy ion fractions, and the distortion of the electric potential is also larger. After the freeze-out of the nucleosynthesis at $T=\mathcal{O}(0.01\,{\rm MeV})$, the dielectric permittivity is nearly constant while the thermal velocity reduces. A slow nucleus results in an almost spherical shape of electric potential as shown in the lower-right panel of figure \ref{fig3}. For $\rho_y=0$, figure \ref{fig3_add} shows the same calculation results with figure \ref{fig3}, in which we can see the apparent difference between forward and backward directions of a moving proton depending on temperatures. In a nutshell, we note that the polarizability of the electric field is intensified by the high velocity of test charge and the deviation of dielectric permittivity. The dynamical screening effect appears by the deformed shape of the electric potential.

\section{Dynamical screening effects on primordial abundances}
In this section, we introduce the formalism of thermonuclear reaction rate including dynamical screened potential. For a two-body reaction between species $1$ and $2$, the general form of the thermal averaged nuclear reaction rate is given as
\begin{eqnarray}
N_A \left\langle \sigma v \right\rangle &=& N_A \int d {\bf v}_1 f({\bf v}_1)\int d{\bf v}_2 f({\bf v}_2) v_r \sigma(E),
\label{reac}
\end{eqnarray}
where $N_A$ is Avogadro's number, ${\bf v}_i$ the velocity of particle $i$, $f({\bf v}_i)$ the normalized velocity distribution function of particle $i$, $v_r=|\bf{v}_1 -{\bf v}_2|$ the relative velocity, and $\sigma(E)$ the cross section for the given reaction depending on the total kinetic energy of $E= \mu v_r^2/2$ with the reduced mass $\mu$ of particles $1$ and $2$ in the center of mass (CM) frame. In the weak screening condition, the transformation of equation \eqref{reac} to CM frame with dynamical screening potential gives the following thermal nuclear reaction rate \cite{1988ApJ...331..565C}:
\begin{eqnarray}
N_A \left\langle \sigma v \right\rangle &=& N_A \sqrt{\frac{8}{\pi \mu T^3}} \int_0^\infty \sigma(E) E f_s(E) e^{-E/T} dE.
\label{eq_rate}
\end{eqnarray}
The enhancement factor by dynamical screening is given as
\begin{eqnarray}
f_s(E) &=& \int_0^\infty \int_0^\pi E^{1/2}_{\rm cm} \pi^{-1/2} T^{-3/2} \sin \theta e^{-E_{\rm cm}/T} e^{W(E,E_{\rm cm},\theta)/T} d\theta dE_{\rm cm},
\label{eq_enhance}
\end{eqnarray}
where $E_{\rm cm}$ is the kinetic energy of the CM and $\theta$ is the angle between the relative velocity and the velocity of the CM. The factor $W(E, E_{\rm cm}, \theta)$ can be transformed from the $W(v_1,v_2,\theta_1,\theta_2)$ defined as
\begin{eqnarray}
W(v_1,v_2,\theta_1,\theta_2) &=& - \frac{Z_1Z_2e^2}{4\pi^2} \int \frac{d {\bf k}}{k^2}  \left[ e^{-i {\bf k} \cdot ({\bf v}_1 - {\bf v}_2)t }  \left( \frac{1}{\epsilon_l({\bf k}, {\bf k} \cdot {\bf v}_1)} -1 \right) \right.  \\[12pt] \nonumber 
&+& \left. e^{-i {\bf k} \cdot ({\bf v}_1 - {\bf v}_2)t }  \left( \frac{1}{\epsilon_l({\bf k}, {\bf k} \cdot {\bf v}_2)} -1 \right) \right].
\label{eq_W}
\end{eqnarray}
We note that this term corresponds to the Debye-H\"uckel potential if we take ${\bf v}_1 = {\bf v}_2 =0$.  Namely, the dynamical screening termed in this paper already reflects the static scheme. Also, unlike the static screening effects, $f_s(E)$ in equation \eqref{eq_enhance} depends not only on the charge number (equation \eqref{eq_W}) but on the dielectric permittivity of plasma related to the thermal velocity.

\begin{figure*}[h]
\centering
\includegraphics[width=7.5cm]{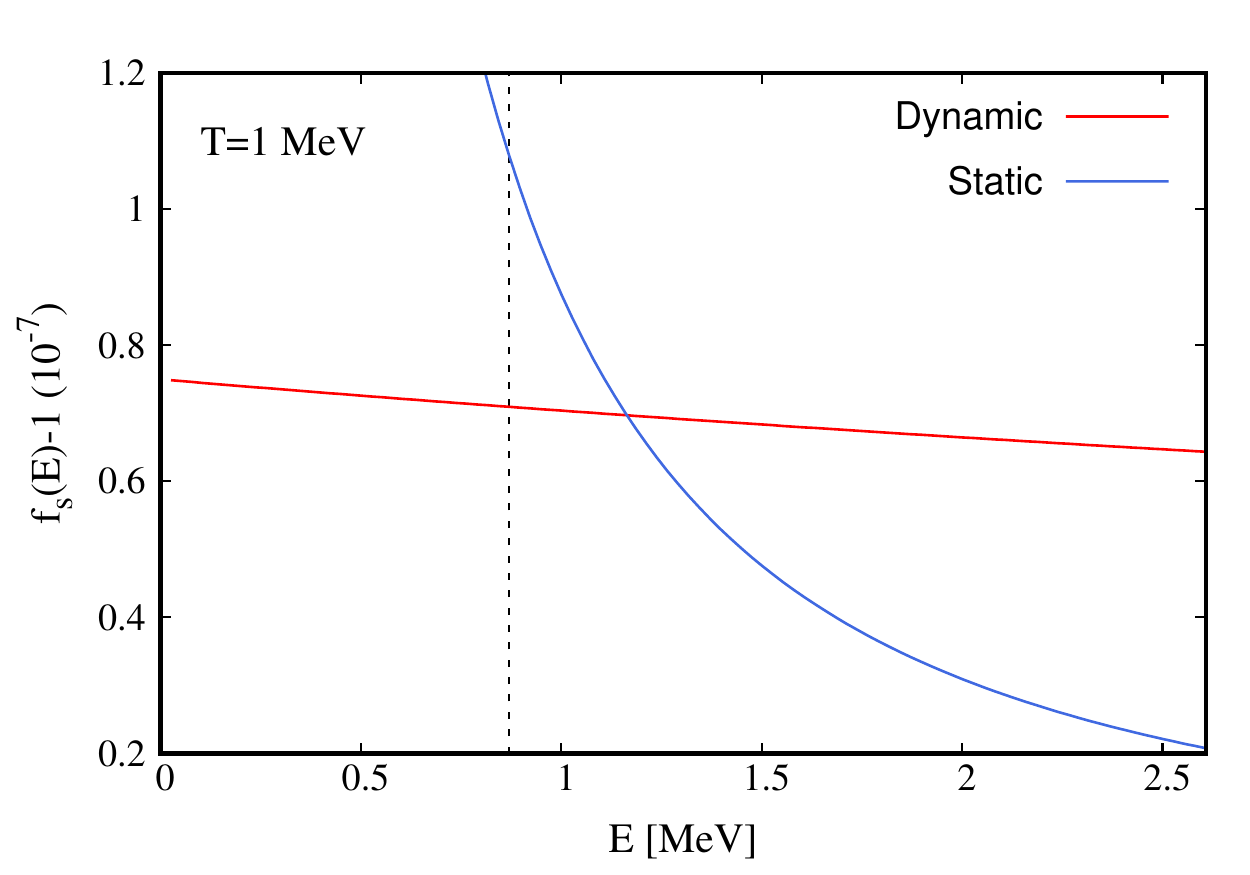}
\includegraphics[width=7.5cm]{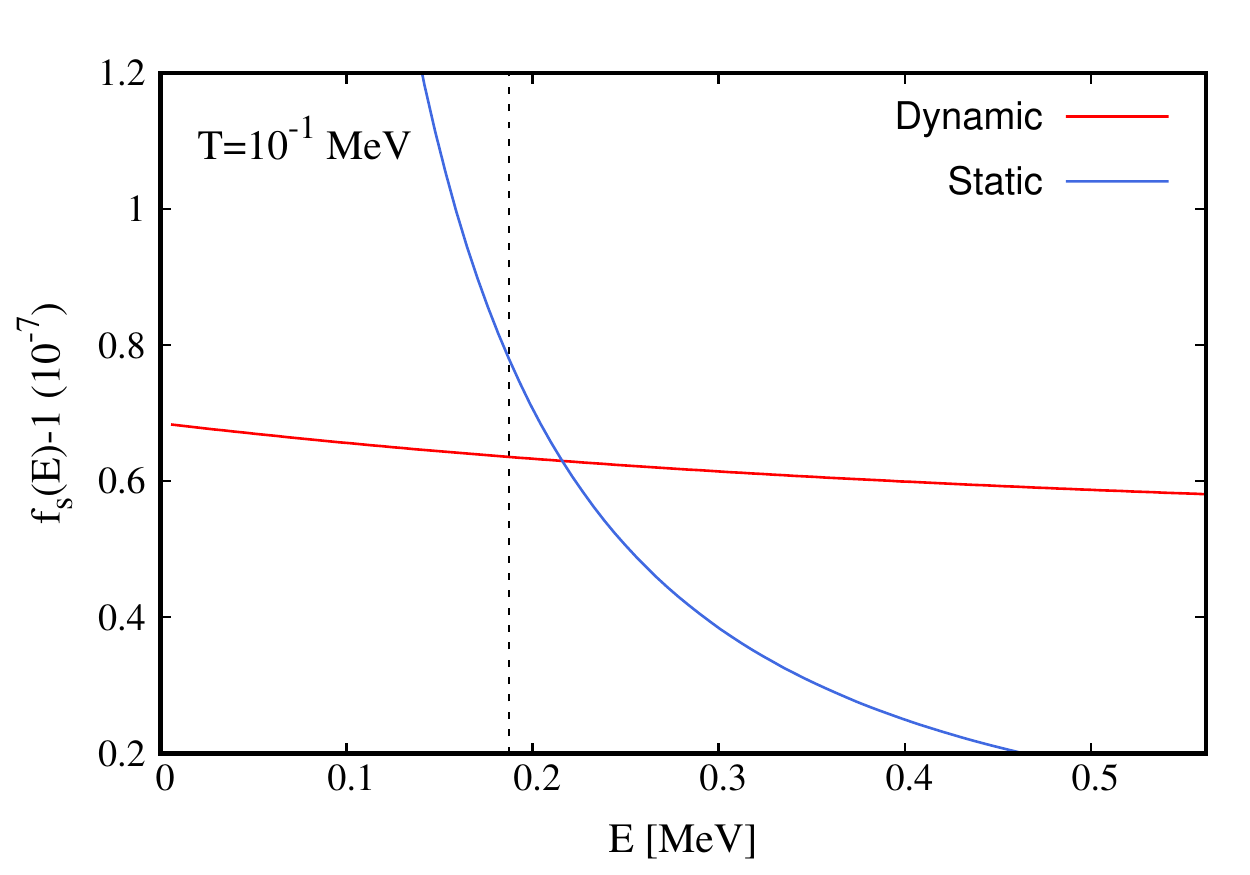}
\includegraphics[width=7.5cm]{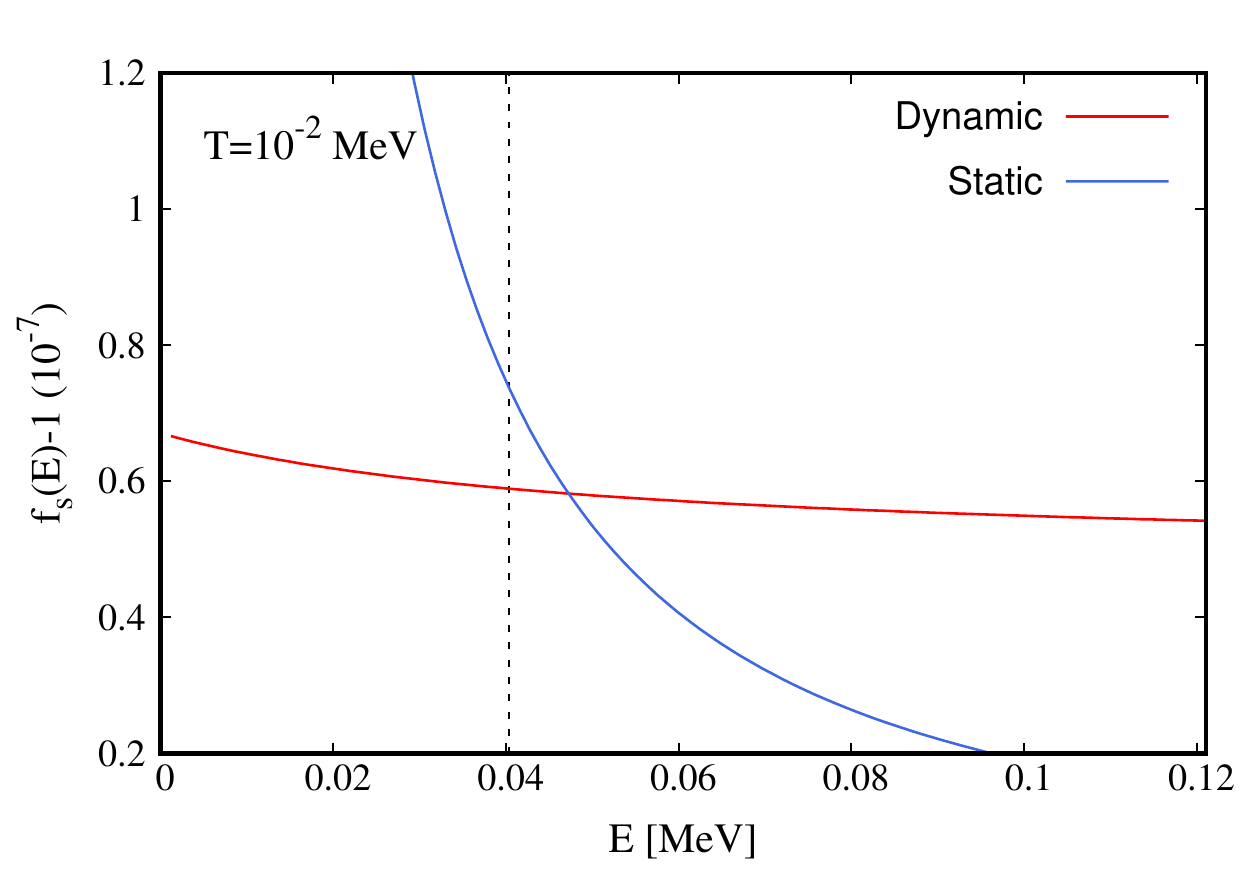}
\includegraphics[width=7.5cm]{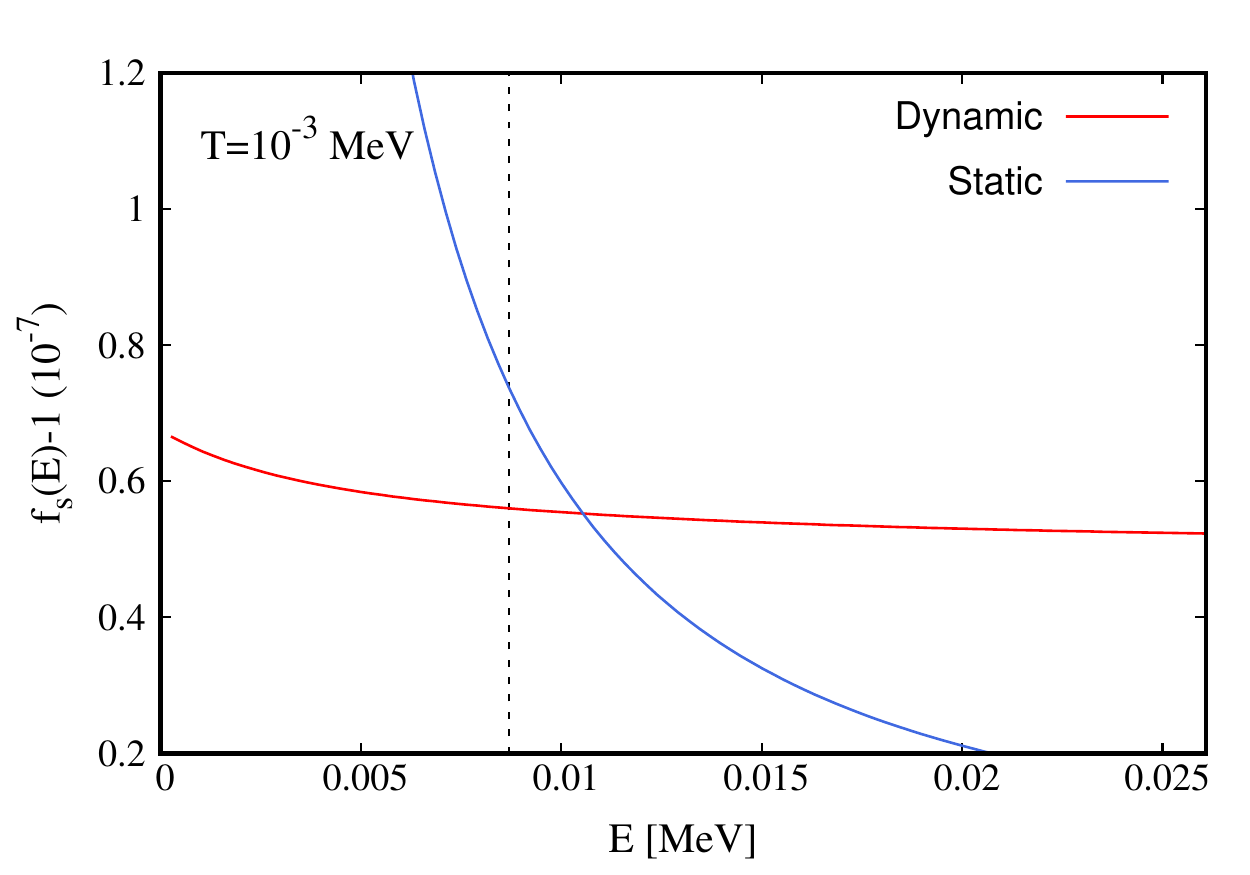}
\caption{Deviation from the unity of the enhancement factor for the reaction of $^2{\rm H}(p,\gamma)^3{\rm He}$ as a function of energy at $T$ = 1\,MeV (upper-left), 0.1\,MeV (upper-right), 0.01\,MeV (lower-left), and 0.001\,MeV (lower-right), respectively. Red and blue lines denote the enhancement factors for dynamical and static screening potentials, respectively. Black vertical dashed lines indicate the Gamow energies at given temperatures.}
\label{fig4}
\end{figure*}

For the reaction of $^2{\rm H}(p,\gamma)^3{\rm He}$, the enhancement factor $f_s(E)$ is shown in figure \ref{fig4} for four temperatures. In general, the screening effect decreases with increasing energy because we can neglect the screening term when the electric potential becomes much lower than the kinetic energy of nuclei. Noteworthy is that the static screening potential is more decreased than the dynamic one in the high energy region. This is the reason why the fast ion forms an intense polarization of potential. On the other hand, in the low energy region, slow ions cannot effectively cause the polarization of the electric field, and the enhancement factor by the dynamical screening effect is reduced more than the static case.

In the integral over energy in the thermonuclear reaction rates (equation \eqref{eq_rate}), at the Gamow energy, the differential rate is maximally impacted. Although the screening energy slightly changes the Gamow peak position, we adopt the standard Gamow energy formula at each given temperature because the change by a shift of the Gamow peak is insignificant. Figure \ref{fig4} shows that the dynamical screening effect is smaller than the static one at the Gamow window. The fact that the enhancement factor is almost constant with respect to energy allows us the following approximation:
\begin{eqnarray}
\left\langle \sigma v \right\rangle_{\rm dynamical} \simeq f_s(E_G) \left\langle \sigma v \right\rangle_{\rm bare},
\label{rate}
\end{eqnarray}
where $E_G$ and $\left\langle \sigma v \right\rangle_{\rm bare}$ stand for the Gamow energy and the bare reaction rate, respectively.

The deuterium abundance is related to the baryon-to-photon ratio observed in CMB and can be compared with astronomical observations of D/H from the analysis of quasar photon spectra \cite{2018ApJ...855..102C}. For evaluation of primordial abundances as well as D destruction in stars, the precise measurements of related reaction cross sections are important. A recent experiment of $^2{\rm H}(p,\gamma)^3{\rm He}$ reaction performed in Laboratory for Underground Nuclear Astrophysics (LUNA) remarkably narrows the uncertainty of the cross section \cite{2020Natur.587..210M}. According to our calculation, the dynamical screening effect changes the thermonuclear reaction rate of $^2{\rm H}(p,\gamma)^3{\rm He}$ by the order of $\lesssim 10^{-7}$, which is allowed within the uncertainty of thermonuclear reaction rates obtained from analysis of LUNA experiments.

\begin{figure}[t]
\centering
\includegraphics[width=11 cm]{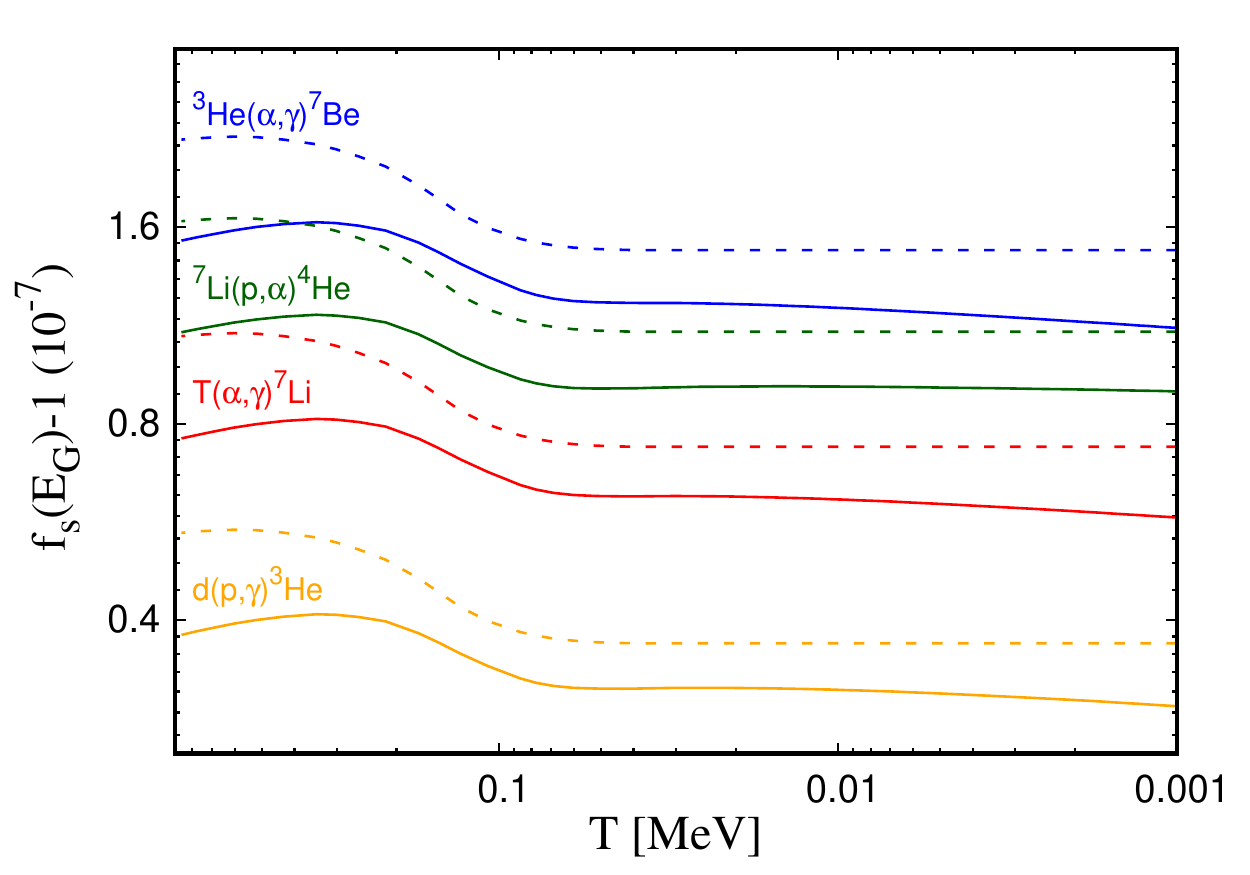}
\caption{Deviation of enhancement factor from unity as a function of cosmic temperature. Solid and dashed lines denote the dynamical and static screening effects, respectively.}
\label{fig5}
\end{figure}

Figure \ref{fig5} shows the enhancement factors for main reactions in BBN by dynamical and static screening effects as a function of cosmic temperature. As shown in figure \ref{fig3}, the Coulomb potential by dynamic nuclei depends on the direction; the potential in forward direction of the moving ion increases, while the one in backward direction decreases. In the colliding system where the interacting particles approach to each other, the Coulomb potential in the forward direction predominantly affects the reaction, which results in the higher Coulomb barrier than the one by the Debye-H\"uckel potential. Therefore, overall temperature region, the dynamical screening enhancement is lower than the static case because the thermonuclear reaction rates are hardly changed. Due to the small change of thermonuclear reaction rates, the effects of dynamical screening on primordial abundances are negligible.

According to reference \cite{Coc:2010zz}, the primordial abundances ($Y_j$) have sensitivities to each main reaction rate in BBN. We can define the sensitivity as follows: 
\begin{eqnarray}
\frac{\partial \ln Y_j}{\partial \ln \left\langle \sigma v \right\rangle}_i = \alpha_{ji},
\end{eqnarray}
where $i$ and $j$ denote the type of reactions and species, respectively. For the small corrections by the screening effects to each reaction rate, the linear approximation leads to the final abundances as follows:
\begin{eqnarray}
Y_j &\simeq& Y_{j, standard} \left( 1 + \sum_i \frac{\Delta \left\langle \sigma v \right\rangle_i}{\left\langle \sigma v \right\rangle_i} \alpha_{ji} \right) \\[12pt]
&=& Y_{j, standard} \left[ 1 + \sum_i \left( \frac{\left\langle \sigma v \right\rangle_{i,dynamical} - \left\langle \sigma v \right\rangle_{i,bare}}{\left\langle \sigma v \right\rangle_{i,bare}} \right) \alpha_{ji} \right] \\[12pt]
& \simeq & Y_{j, standard} \left[ 1 + \sum_i \left( \frac{f_s(E_G)_i \left\langle \sigma v \right\rangle_{i,bare} - \left\langle \sigma v \right\rangle_{i, bare}}{\left\langle \sigma v \right\rangle_{i,bare}} \right) \alpha_{ji} \right] \\[12pt]
& = & Y_{j, standard} \left[ 1 + \sum_i \left( f_s(E_G)_i - 1 \right) \alpha_{ji} \right], 
\end{eqnarray}
where $Y_{j,standard}$ denotes the final abundances of standard BBN calculation with bare reaction rates. For the sensitivity $\alpha_{ji}$, we adopt table 1 in reference \cite{Coc:2010zz}. Although the enhancement factor depends on temperature, it behaves like a constant as shown in figure \ref{fig5}. Hence, we choose the averaged value of the enhancement factor for estimating the final abundances.

Table \ref{table1-1} shows the fractional change of final abundances between standard and dynamical (static) screening cases. The screening effects, which increase charged particle reaction rates, activate the nucleosynthesis. The increase in deuterium destruction via $^2$H(p,$\gamma$)$^3$He, $^2$H(d,n)$^3$He, and $^2$H(d,p)$^3$H decreases D/H, which also leads to the reduction of $^3$He/H. Besides, the decrease in D/H results in smaller neutron number, slowing down the neutron induced reaction rates like $^1$H(n,$\gamma)^2$H and $^7$Be(n,p)$^7$Li. Since the latter reaction is the main destruction process of $^7$Be, the suppression of $^7$Be(n,p)$^7$Li leaves more $^7$Be which finally decays to more $^7$Li by electron capture. Note that the $^4$He abundance does not have any appreciable difference because it depends strongly on only the $n \leftrightarrow p$ weak reactions at earlier epoch \cite{Coc:2010zz} which are almost free from the dynamical screening effects in the present study. For the static case, the above effects are larger than the dynamical screening case because the enhancement factor by the static screening is slightly larger than one in the static case. 
\begin{table}[h]
    \centering
    \begin{tabular}{c | c c c c} \hline\hline
    Ratio                            &    $Y_p$   &     D/H    &  $^3$He/H  &  $^7$Li/H      \\ \hline\hline
    $R_{dyn/stand}\ \    (10^{-8})$  &     0      &    -4.29   &   -3.94     &    11.7         \\ \hline
    $R_{stat/stand}\ \   (10^{-8})$  &     0      &    -5.62   &   -5.14     &    15.5         \\ \hline
    $R_{dyn/stat}\ \ \ \ (10^{-8})$  &     -      &     1.32   &    1.21     &    -3.77         \\ \hline
    \end{tabular}
    \caption{The fractional differences of final abundances between standard and dynamical (static) screening cases. Each ratio is defined as follows: $R_{dyn/stand}=(Y_{j,dynamical}-Y_{j,standard})/Y_{j,standard}$, $R_{stat/standard}=(Y_{j,static}-Y_{j,standard})/Y_{j,standard}$, and $R_{dyn/stat}=(Y_{j,dynamical}-Y_{j,static})/Y_{j,static}$.}
    \label{table1-1}
\end{table}

\section{Discussion and conclusions}
 Although the early universe has high enough temperatures to produce feasible velocities of ions, low particle densities cannot effectively change the dielectric permittivity. This is contrary to the results in the solar condition. For several important reactions in solar fusion, we show updated enhancement factors by the dynamical screening effects in table \ref{table1}. We adopt the same method used in reference \cite{1988ApJ...331..565C}, but the different electron density derived by the Fermi-Dirac distribution with the given temperature. This indicates a $\sim$10\% increase of the enhancement factor in the third column and a reduction by the dynamical screening in the fourth column. Although opinions vary in the literature on the adopted method and result \cite{1997RvMP...69..411B, 1998ApJ...496..503G, 1999PhR...311...99S, 2000ApJ...535..473O}, if the dynamical screening effects are visible under the solar condition as suggested in references \cite{2010Ap&SS.328..153M, 2017PhRvD..95k6002Y} as well as \cite{1988ApJ...331..565C}, those effects leave several issues worth discussing for related plasma properties in other astrophysical environments.

\begin{table}[h]
    \centering
    \begin{tabular}{cccc} \hline\hline
    reaction      & $E_G/T$ & $f_s(E_G)$ &  $f_s (E_G)/f_{en} (E_G)$\\ \hline\hline
    $p$-$p$         & 4.5689  & 1.0262     &  0.9889        \\ \hline
    $^3$He-$^3$He & 16.5938 & 1.1078     &  0.9591        \\ \hline
    $^3$He-$^4$He & 17.3375 & 1.1087     &  0.9599        \\ \hline
    $p$-$^7$Be    & 13.8710 & 1.1187     &  0.9686        \\ \hline
    $p$-$^{14}$N    & 20.5473 & 1.2242     &  0.9514        \\ \hline
    \end{tabular}
    \caption{The comparison of enhancement factors between dynamical and static screening effects for several important reactions in the solar environment. The $f_s$ (equation \eqref{eq_enhance}) and $f_{en}$ (equation \eqref{salpeter}) stand for the enhancement factors by dynamical and static charges, respectively. For this calculation, we adopt the following conditions: mass density $\rho=1.6 \times 10^2\,{\rm g/cm^{3}}$; temperature $T=1.25 \times 10^7\,{\rm K}$; mass fraction of proton $X=0.7$; and mass fraction of $^4$He $Y_p=0.3$.}
    \label{table1}
\end{table}

\begin{figure}[t]
\centering
\includegraphics[width=11 cm]{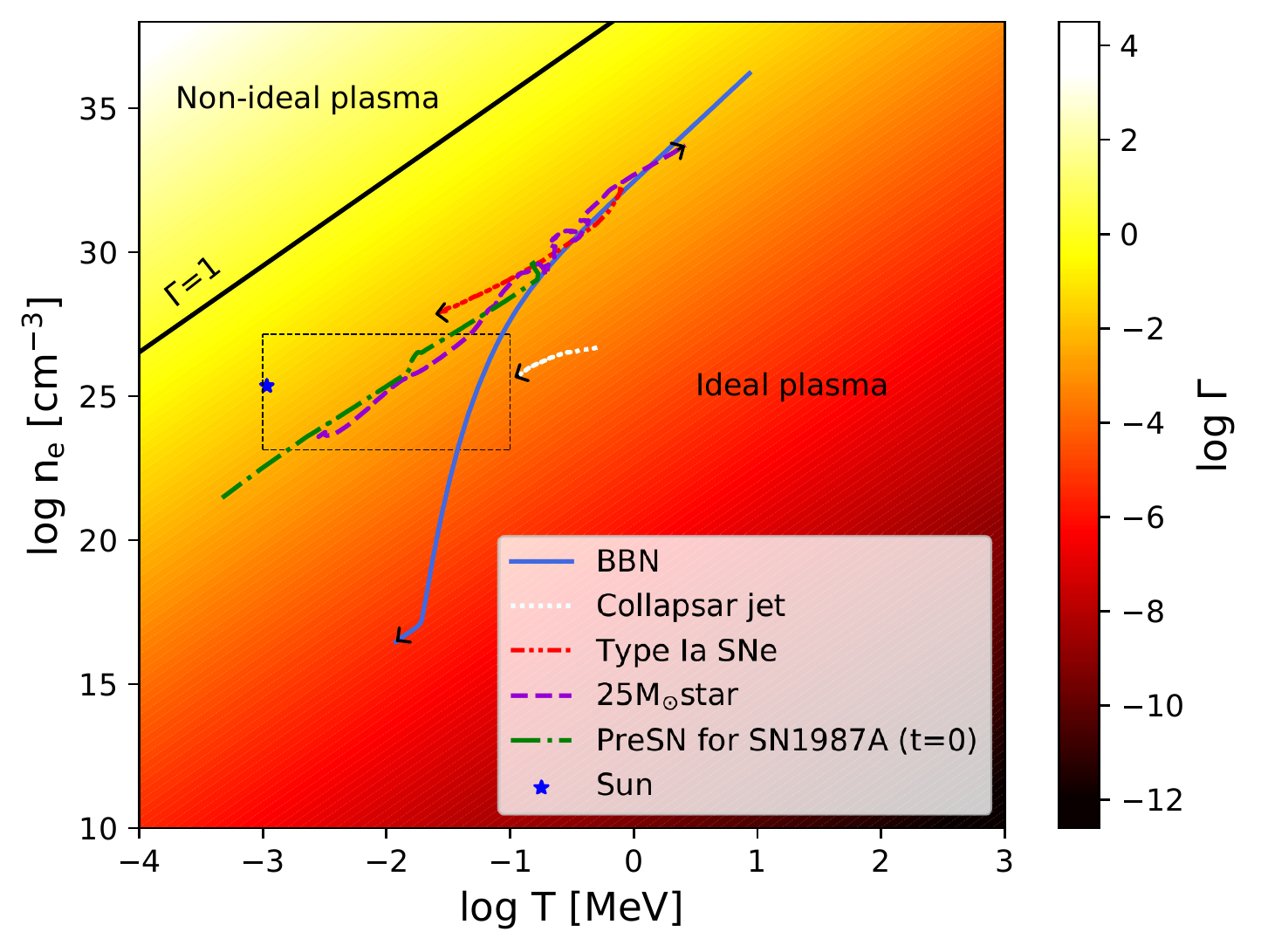}
\caption{Classical plasma parameter $\Gamma$ $= n_e^{1/3} (Ze)^2 / T$ and trajectories of BBN (blue-solid), ejected collapsar jet (white-dotted), the center of Type Ia supernovae (red-dashed-double dotted), the center of $25\,{\rm M}_{\odot}$ star (purple-dashed), the initial profile of preSN for SN1987A (green-dashed-single dotted), and the sun (blue star mark) as a function of $T$ and $n_e$. Each arrow indicates the direction of the evolution. The black-solid line indicates the contour line of $\Gamma=1$ and the open box corresponds to the region of parameter space discussed in figure \ref{fig8}.}
\label{fig7}
\end{figure}

First, a plasma state can be characterized with the plasma coupling parameter, classically defined as $\Gamma = n_e^{1/3} (Ze)^2 / T$, which means a ratio of mean Coulomb energy to averaged kinetic energy. (For a degenerate system, the quantum plasma parameter depending on the Fermi energy is required to classify the ideality of plasma state, but we consider the only classical regime in this paper.) Figure \ref{fig7} shows the classical plasma coupling parameter and trajectories of denoted astrophysical environments on the parameter plane of temperature and electron number density $n_e$. For the density evolution, we adopt trajectories of the BBN calculated by the updated code from \cite{1992STIN...9225163K,  1993ApJS...85..219S}, the 25 solar mass (${\rm M}_\odot$) star from \cite{2011ApJS..192....3P, 2015ApJS..220...15P, 2018ApJS..234...34P, 2019ApJS..243...10P}, the model of collapsar jet with a 5 degree ejection angle in \cite{2015A&A...582A..34N}, Type Ia Supernovae (SNe) for W7 model in \cite{1984ApJ...286..644N, 2020ApJ...904...29M}, the initial preSN for SN1987A model from \cite{1984ApJ...286..644N, 1990ApJ...360..242S} and the condition of the sun used in table \ref{table1}. According to our investigation, such astrophysical environments are located in the ideal or weakly non-ideal plasma region---justifying the weak screening condition. On the other hand, for more extreme conditions satisfying $\Gamma \gtrsim 1$, it may require the physics beyond the weak screening assumption using the quantum electrodynamics in the finite temperature medium involving many-body interactions.

\begin{figure*}[t]
\centering
\includegraphics[width=7.5cm]{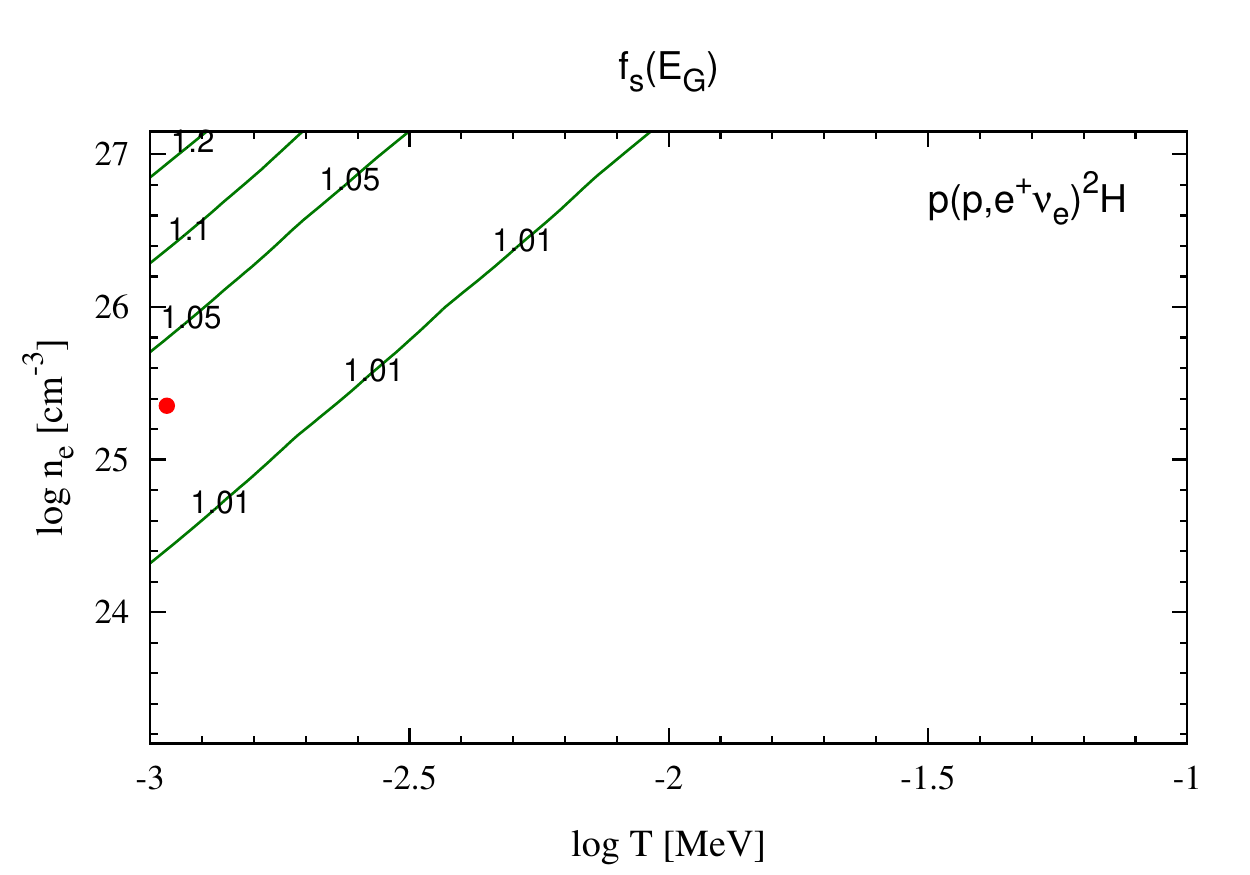}
\includegraphics[width=7.5cm]{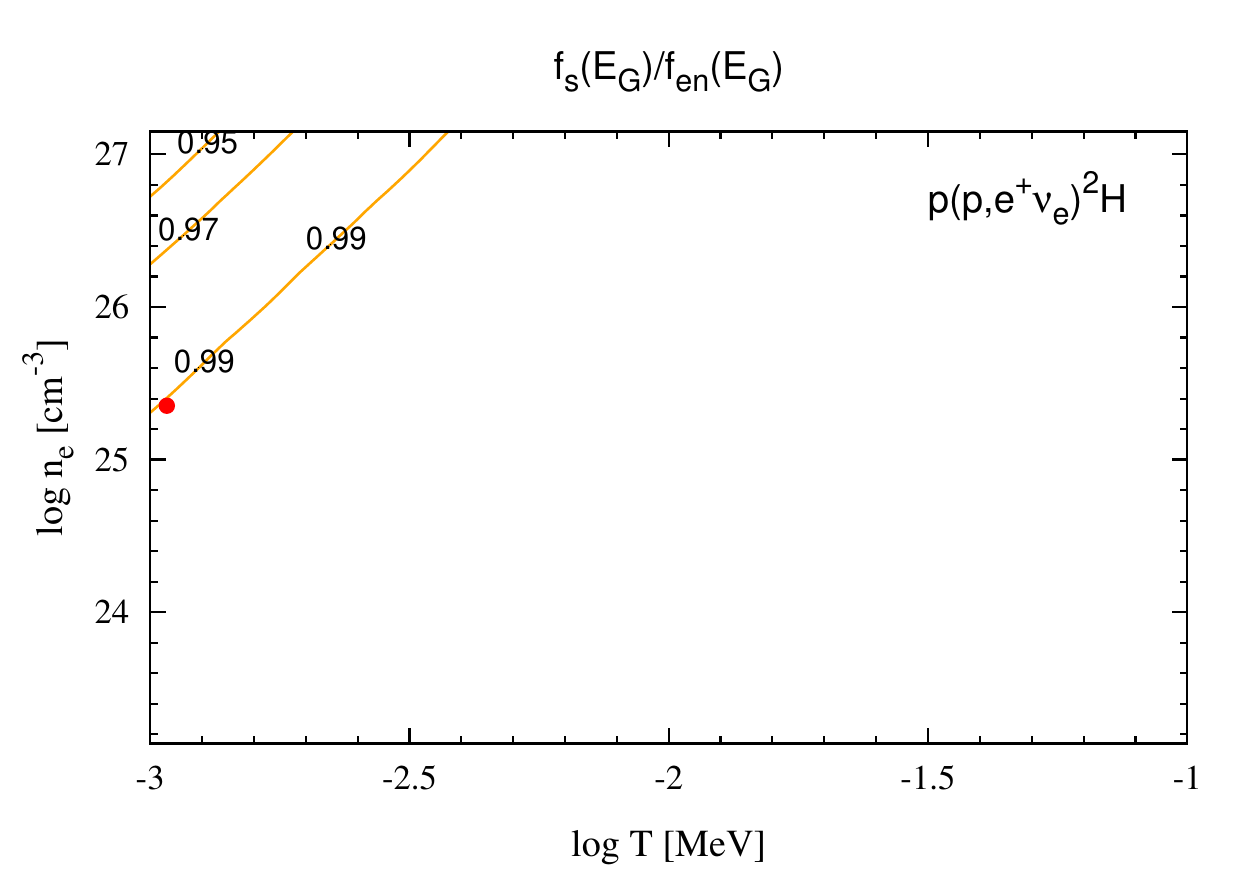}
\includegraphics[width=7.5cm]{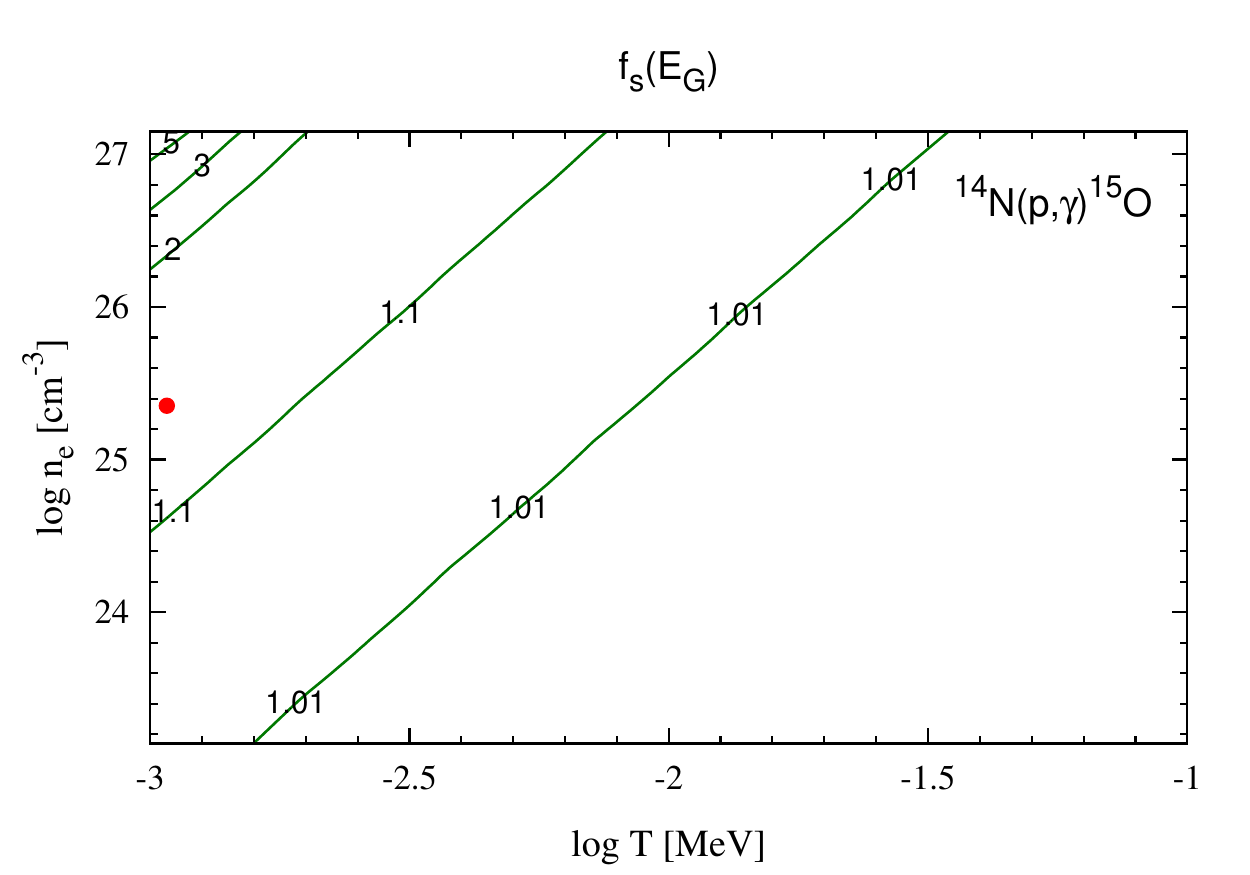}
\includegraphics[width=7.5cm]{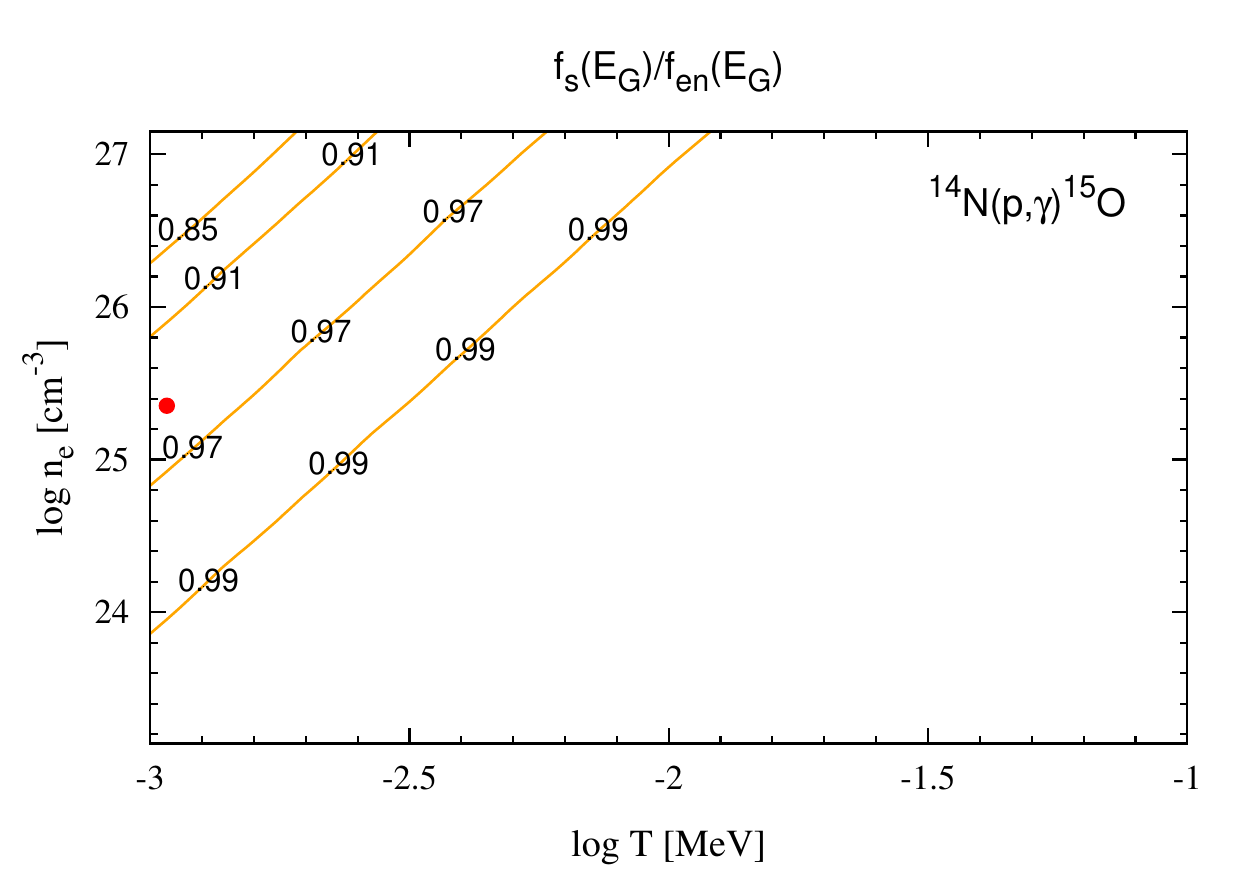}
\caption{Enhancement factor by dynamical screening (left panels) and ratio of static to dynamical enhancement factor, i.e., $f_{s} (E_G)/f_{en}(E_G)$ (right panels) as a function of $T$ and $\rho$ in the region of the open box in figure \ref{fig7}. The upper and lower panels show the results for the reactions of $p(p,e^+\nu_e)^2{\rm H}$ and $^{14}{\rm N}(p,\gamma)^{15}{\rm O}$, respectively. We adopt the solar condition denoted by the red point and mass fractions of nuclei used in table \ref{table1}.}
\label{fig8}
\end{figure*}

Second, even in the weakly coupled plasma region, the left panel in figure \ref{fig8} implies that the enhancement factor by the dynamical screening effects can be significant (See also open box in figure \ref{fig7}.). This alludes to that the thermonuclear reaction rates in other astrophysical environments can also be affected by the dynamical screening effects under the weak screening condition. For example, in the solar condition, contour lines in figure \ref{fig8} show the remarkable enhancements for reaction rates of $p(p,e^+\nu_e)^2{\rm H}$ and $^{14}{\rm N}(p,\gamma)^{15}{\rm O}$ (See also table \ref{table1}.). Both reactions are important in that the former reaction triggers the solar fusion known as the $pp$ chain and the latter is one of the CNO cycle reactions both of which operate for He synthesis in main-sequence stars. 
In particular, relevant reactions of CNO cycle dominating the solar energy generation significantly impacts on not only the evolution of the solar density profiles but the CNO neutrinos recently detected \cite{2020Natur.587..577B}. Although the dynamical screening effects are lower than the static ones as shown in right panels in figure \ref{fig8}, the enhancement of reaction rates increases with charge numbers of interacting nuclei similar to the static case. This means that the dynamical screening effects on the CNO cycle are more effective than that on the $pp$ chain, which could provide the correction of the CNO cycle for the solar evolution as well as CNO neutrino detection. Furthermore, we expect the conceivable changes of electron capture rates \cite{2020PhRvD.101h3010L, 2020ApJ...904...29M, Itoh:2002ps} and other thermonuclear reaction rates in various astrophysical environments can result in a change of relevant nuclear processes. Such a change would play significant roles in the theoretical predictions of neutrino signal or the rate of gravitational waves emission from the stellar objects as well as the elemental yields by the nucleosynthesis in the cosmos.

Third, the correlation energy from the electromagnetic fluctuation can be produced as the plasma parameter increases \cite{2001PhPl....8.2454O}. Such effects on the thermodynamic quantities including the energy density and pressure in the early universe were investigated \cite{Itoh et al.(1997)}. The correlation in microstates is also one of the motivations to invent the non-extensive statistics, known as Tsallis statistics \cite{1988JSP....52..479T}. By the non-extensive statistics, we can consider a non-Maxwellian distribution in dense astrophysical environments to properly describe the deviation from the standard thermodynamics. For the BBN calculation, the effects of non-extensive statistics are investigated \cite{Bertulani:2012sv, Hou:2017uap}. However, the latest study shows that those results for thermonuclear reactions are not consistent with observations \cite{2019PhRvD..99d3505K}. Also, according to our result shown in figure \ref{fig8}, we expect that the small value of plasma coupling parameter in the BBN epoch would not allow the strongly correlated system. This means that the nuclear distribution function rarely deviates from the equilibrium by the electromagnetic correlation during the BBN epoch. Using the Boltzmann equation and Monte-Carlo simulation, respectively, references \cite{McDermott:2018uqm} and \cite{Sasankan:2019oee} show that the nuclear distribution function during the BBN epoch is close to the equilibrium, i.e., Maxwell-Boltzmann distribution. However, the possibility of the non-Maxwellian distribution for the thermonuclear reaction rates remains in other astrophysical sites.

Lastly, we address the effects of the transverse mode of permittivity in astrophysical nuclear processes. According to reference \cite{1997PhRvL..79.2628O}, by the fluctuation-dissipation theorem, the transverse mode of dielectric permittivity affects the electromagnetic fluctuation \cite{2020PhPl...27b2106B}, by which the low-frequency region of electromagnetic spectra deviates from the black body form. In the BBN calculation, the deviation of photon spectra from the black body would affect the photo-disintegration rates as well as the number or energy density of photons. Reference \cite{2018arXiv181209472J} shows that the deviation of the photon spectra can be a solution to the primordial lithium problem, although they adopt an increase of high energy tail with the help of Tsallis statistics for the Bose-Einstein distribution. Hence, whether the transverse mode of dielectric permittivity can affect the BBN or not should be investigated in the future.

In conclusion, we present the effects of dynamical screening on BBN using the test charge method. In the early phase of the BBN, a plasma is dominated by relativistic $e^-$ and $e^+$. The equality between $e^-$ and $e^+$ number densities leads to canceling out the net current, and the dielectric permittivity is not significantly changed from that of the free space. After that, during the BBN in the adiabatic expansion of the universe, the difference in number density between $e^-$ and $e^+$ becomes relatively larger at a later time when the annihilation proceeds. Then, the dielectric permittivity deviates from the unity. However, the drastically dropped electron number density strongly suppresses the dynamical screening effects. Consequently, our result shows that the dynamical screening effects do not affect the primordial abundances. Also, for other astrophysical nuclear processes, we discuss the plasma properties not considered in our present calculation: (1) a plasma coupling parameter, (2) the dynamical screening effects on CNO neutrinos, (3) the correlation energy and non-extensive statistics, and (4) the transverse mode of the dielectric permittivity. We expect that such a proper consideration of the corrections for astrophysical nuclear processes would advance further the understanding of the origin of elements in the future.

\section*{Acknowledgments}
D.J. and E.H. are grateful to H. Ko and Y. Kwon for useful discussions and information. The work of D.J. and C.M.R. is supported by Institute for Basic Science under IBS-R012-D1. E.H, K.P and M.K.C are supported by the National Research Foundation of Korea (Grant Nos. NRF-2013M7A1A1075764, NRF-2020R1A2C3006177, and NRF-2021R1A6A1A03043957). K.P is also supported by NRF-2021R1I1A1A01057517. The work of M.K. was supported by NSFC Research Fund for International Young Scientists (11850410441). T.K. is supported in part by Grants-in-Aid for Scientific Research of JSPS (20K03958, 17K05459) and T.M. is supported by JSPS (19K03833). A.B.B. is supported in part by the U.S. National Science Foundation grant No. PHY-2108339 and acknowledges support from the NAOJ visiting professor program.



\begin{thebibliography}{99}

\bibitem{1954AuJPh...7..373S}  E.~E.~Salpeter 
\emph{Electron Screening and Thermonuclear Reactions}, 
\emph{Australian Journal of Physics} {\bf 7} (1954) 373.

\bibitem{2013CoPP...53..397P}  A.~Y.~Potekhin and G.~Chabrier, 
\emph{Electron Screening Effect on Stellar Thermonuclear Fusion}, 
\emph{Contributions to Plasma Physics} {\bf 53} (2013) 397 [arXiv:1310.3162].


\bibitem{2007A&A...463..261L} M.~Q.~Liu, J.~Zhang and Z.~Q.~Luo, 
\emph{Screening effect on electron capture in presupernova stars}, 
\emph{Astron. Astrophys.} {\bf 463} (2007) 261.


\bibitem{2009MNRAS.400..815L} M.-Q.~Liu, Y.-F.~Yuan and J.~Zhang, 
\emph{Effect of electron screening on the collapsing process of core-collapse supernovae}, 
\emph{Mon. Not. Roy. Astron. Soc.} {\bf 400} (2009) 815.


\bibitem{2011PhRvC..83a8801W} B.~Wang, C.~A.~Bertulani and A.~B.~Balantekin, 
\emph{Electron screening and its effects on big-bang nucleosynthesis}, 
\emph{Phys. Rev. C} {\bf 83} (2011) 018801 [arXiv:1010.1565].


\bibitem{2016PhRvC..93d5804F} M.~A.~Famiano, A.~B.~Balantekin and T.~Kajino, 
\emph{Low-lying resonances and relativistic screening in Big Bang nucleosynthesis}, 
\emph{Phys. Rev. C} {\bf 93} (2016) 045804 [arXiv:1603.03137].


\bibitem{2020PhRvD.101h3010L} Y.~Luo et al., 
\emph{Screening corrections to electron capture rates and resulting constraints on primordial magnetic fields},
\emph{Phys. Rev. D} {\bf 101} (2020) 083010 [arXiv:2002.08636].


\bibitem{1981JPlPh..25..225W} C.~L.~Wang, G.~Joyce and D.~R.~Nicholson, 
\emph{Debye shielding of a moving test charge in plasma}, 
\emph{Journal of Plasma Physics} {\bf 25} (1981) 225.


\bibitem{1993JETP...77..910T} {\'E}.~{\'E}~Trofimovich and V.~P.~Krainov, 
\emph{Shielding of a moving charge in a Maxwellian plasma}, 
\emph{Soviet Journal of Experimental and Theoretical Physics} {\bf 77} (1993) 910.


\bibitem{1977ApJ...212..513M} H.~E.~Mitler, 
\emph{Thermonuclear ion-electron screening at all densities. I. Static solution}, 
\emph{Astrophys. J.} {\bf 212} (1977) 513.


\bibitem{1988ApJ...331..565C} C.~Carraro, A.~Schafer and S.~E.~Koonin, 
\emph{Dynamic Screening of Thermonuclear Reactions}, 
\emph{Astrophys. J.} {\bf 331} (1988) 565.


\bibitem{2016RvMP...88c0502M} A.~B.~McDonald, 
\emph{Nobel Lecture: The Sudbury Neutrino Observatory: Observation of flavor change for solar neutrinos},
\emph{Rev. Mod. Phys.} {\bf 88} (2016) 030502.


\bibitem{2016RvMP...88c0501K} T.~Kajita, 
\emph{Discovery of atmospheric neutrino oscillations}, 
\emph{Rev. Mod. Phys.} {\bf 88} (2016) 030501.


\bibitem{1998ApJ...496..503G} A.~V.~Gruzinov, 
\emph{Dynamic Screening and Thermonuclear Reaction Rates}, 
\emph{Astrophys. J.} {\bf 496} (1998) 503 [arXiv:astro-ph/9702064].


\bibitem{2002A&A...383..291B} J.~N.~Bahcall, L.~S.~Brown, A.~Gruzinov and R.~F.~Sawyer, 
\emph{The Salpeter plasma correction for solar fusion reactions}, 
\emph{Astron. Astrophys.} {\bf 383} (2002) 291 [arXiv:astro-ph/0010055].


\bibitem{1997RvMP...69..411B} L.~S.~Brown and R.~F.~Sawyer, 
\emph{Nuclear reaction rates in a plasma}, 
\emph{Rev. Mod. Phys.} {\bf 69} (1997) 411 [arXiv:astro-ph/9610256].


\bibitem{2000ApJ...535..473O} M.~Opher and R.~Opher, 
\emph{Dynamic Screening in Thermonuclear Reactions}, 
\emph{Astrophys. J.} {\bf 535} (2000) 473 [arXiv:astro-ph/9908218].


\bibitem{2010Ap&SS.328..153M} K.~Mussack and W.~D{\"a}ppen, 
\emph{Dynamic screening in solar and stellar nuclear reactions}, 
\emph{Astrophysics and Space Science} {\bf 328} (2010) 153 [arXiv:0909.2646].


\bibitem{2011ApJ...729...96M} K.~Mussack and W.~D{\"a}ppen, 
\emph{Dynamic screening correction for solar p-p reaction rates}, 
\emph{Astrophys. J.} {\bf 739} (2011) 96 [arXiv:1102.5073].


\bibitem{1999PhR...311...99S} G.~Shaviv and N.~J.~Shaviv, 
\emph{Is there a dynamic effect in the screening of nuclear reactions in stellar plasmas?}, 
\emph{Phys. Rep.} {\bf 311} (1999) 99. 

\bibitem{2000A&A...356L..57T} V.~N.~Tsytovich, 
\emph{Suppression of thermonuclear reactions in dense plasmas instead of Salpeter's enhancement}, 
\emph{Astron. Astrophys.} {\bf 356} (2000) L57-L61.


\bibitem{2017PhRvD..95k6002Y} X.~Yao, T.~Mehen and B.~M\"uller, 
\emph{Dynamical screening of \ensuremath{\alpha}-\ensuremath{\alpha} resonant scattering and thermal nuclear scattering rate in a plasma}, 
\emph{Phys. Rev. D} {\bf 95} (2017) 116002 [arXiv:1609.00383].


\bibitem{1981phki.book.....L} E.~M.~Lifshitz and L.~P.~Pitaevskii, 
\emph{Physical kinetics}, 
Oxford: Pergamon Press, 1981.


\bibitem{2018PhR...754....1P} C.~Pitrou, A.~Coc, J.~P.~Uzan and E.~Vangioni,
\emph{Precision big bang nucleosynthesis with improved Helium-4 predictions}, 
\emph{Phys. Rep.} {\bf 754} (2018) 1 [arXiv:1801.08023].


\bibitem{1992STIN...9225163K} L.~Kawano,
\emph{Let's go: Early universe. 2. Primordial nucleosynthesis: The Computer way},
FERMILAB-PUB-92-004-A (1992).



\bibitem{1993ApJS...85..219S} M.~S.~Smith, L.~H.~Kawano and R.~A.~Malaney,
\emph{Experimental, Computational, and Observational Analysis of Primordial Nucleosynthesis}, 
\emph{Astrophys. J. Suppl. } {\bf 85} (1993) 219. 


\bibitem{2004ADNDT..88..203D} P.~Descouvemont, A.~Adahchour, C.~Angulo, A.~Coc and E.~Vangioni-Flam,
\emph{Compilation and R-matrix analysis of Big Bang nuclear reaction rates},
\emph{Atom. Data Nucl. Data Tabl.} {\bf 88} (2004) 203 [arXiv:astro-ph/0407101].


\bibitem{2016ApJ...831..107I} C.~Iliadis, K.~Anderson, A.~Coc, F.~Timmes and S.~Starrfield,
\emph{Bayesian Estimation of Thermonuclear Reaction Rates}, 
\emph{Astrophys. J.} {\bf 831} (2016) 107 [arXiv:1608.05853].  



\bibitem{2018PhRvD..98c0001T} M.~Tanabashi \textit{et al.} [Particle Data Group],
\emph{Review of Particle Physics},
\emph{Phys. Rev. D} {\bf 98} (2018) no.3 030001.


\bibitem{2016A&A...594A..13P} PP.~A.~R.~Ade \textit{et al.} [Planck],
\emph{Planck 2015 results. XIII. Cosmological parameters},
\emph{Astron. Astrophys.} {\bf 594} (2016) A13 [arXiv:1502.01589].


\bibitem{2018ApJ...855..102C} R.~J.~Cooke, M.~Pettini and C.~C.~Steidel,
\emph{One Percent Determination of the Primordial Deuterium Abundance},
\emph{Astrophys. J.} {\bf 855} (2018) no.2 102 [arXiv:1710.11129].


\bibitem{2020Natur.587..210M}V.~Mossa, K.~St\"ockel, F.~Cavanna, F.~Ferraro, M.~Aliotta, F.~Barile, D.~Bemmerer, A.~Best, A.~Boeltzig and C.~Broggini, \textit{et al.},
\emph{The baryon density of the Universe from an improved rate of deuterium burning}, 
\emph{Nature} \textbf{587} (2020) no.7833 210.  

\bibitem{Coc:2010zz}A.~Coc and E.~Vangioni,
\emph{Big-Bang nucleosynthesis with updated nuclear data},
\emph{J. Phys. Conf. Ser.} \textbf{202} (2010) 012001. 


\bibitem{2005NuPhB.729..221M} J.~Birrell, C.~T.~Yang and J.~Rafelski, 
\emph{Relic Neutrino Freeze-out: Dependence on Natural Constants},
\emph{Nucl. Phys. B} {\bf 890} (2014) 481-517 [arXiv:1406.1759].


\bibitem{1996RPPh...59.1493S} S.~Sarkar,
\emph{Big bang nucleosynthesis and physics beyond thestandard model}, 
\emph{Reports on Progress in Physics} {\bf 59} (1996) 1493 [arXiv:hep-ph/9602260].


\bibitem{2011ApJS..192....3P} B.~Paxton, L.~Bildsten, A.~Dotter, F.~Herwig, P.~Lesaffre and F.~Timmes,
\emph{Modules for experiments in stellar astrophysics (MESA)}, 
\emph{Astrophys. J. Suppl.} {\bf 192} (2011) 3 [arXiv:1009.1622].


\bibitem{2015ApJS..220...15P} B.~Paxton, P.~Marchant et al.,
\emph{Modules for experiments in stellar astrophysics (MESA): Binaries, pulsations, and explosions},
\emph{Astrophys. J. Suppl.} {\bf 220} (2015) no.1 15 [arXiv:1506.03146].


\bibitem{2018ApJS..234...34P} B.~Paxton et al.,
\emph{Modules for experiments in stellar astrophysics (MESA): Convective boundaries, element diffusion, and massive star explosions}, 
\emph{Astrophys. J. Suppl.} {\bf 234} (2018) 34 [arXiv:1710.08424].


\bibitem{2019ApJS..243...10P} B.~Paxton et al.,
\emph{Modules for Experiments in Stellar Astrophysics (MESA): Pulsating Variable Stars, Rotation, Convective Boundaries, and Energy Conservation}, 
\emph{Astrophys. J. Suppl.} {\bf 243} (2019) 10 [arXiv:1903.01426].


\bibitem{2015A&A...582A..34N} K.~Nakamura, S.~Harikae, T.~Kajino and G.~J.~Mathews,
\emph{r-process nucleosynthesis in the MHD+neutrino-heated collapsar jet}, 
\emph{Astron. Astrophys.} {\bf 582} (2015) 9.


\bibitem{1984ApJ...286..644N} K.~Nomoto, F.-K.~Thielemann and K.~Yokoi, 1984, apj, 286, 644.
\emph{Accreting white dwarf models for type I supernovae. III. Carbon deflagration supernovae}, 
\emph{Astrophys. J.} {\bf 286} (1984) 644.


\bibitem{2020ApJ...904...29M} K.~Mori, T.~Suzuki, M.~Honma, M.~A.~Famiano, T.~Kajino, M.~Kusakabe and A.~B.~Balantekin,
\emph{Screening effects on electron capture rates and type Ia supernova nucleosynthesis}, 
\emph{Astrophys. J.} {\bf 904} (2020) no.1 29 [arXiv:2009.10925].


\bibitem{1990ApJ...360..242S} T.~Shigeyama and K.~Nomoto,
\emph{Theoretical light curve of SN 1987A and mixing of hydrogen and nickel in the ejecta}, 
\emph{Astrophys. J.} {\bf 360} (1990) 242.


\bibitem{2020Natur.587..577B} M.~Agostini et al. [BOREXINO],
\emph{Experimental evidence of neutrinos produced in the CNO fusion cycle in the Sun}, 
\emph{Nature} {\bf 587} (2020) 577 [arXiv:2006.15115].


\bibitem{Itoh:2002ps} N.~Itoh, N.~Tomizawa, M.~Tamamura, S.~Wanajo and S.~Nozawa,
\emph{Screening corrections to the electron capture rates in dense stars by the relativistically degenerate electron liquid},
\emph{Astrophys. J.} \textbf{579} (2002) 380-385 
[arXiv:astro-ph/0207132 [astro-ph]].


\bibitem{2001PhPl....8.2454O} M.~Opher, L.~O.~Silva, D.~E.~Dauger, V.~K.~Decyk and J.~M.~Dawson,
\emph{Nuclear reaction rates and energy in stellar plasmas : the effect of highly damped modes}, 
\emph{Phys. Plasmas} {\bf 8} (2001) 2454 [arXiv:astro-ph/0105153].


\bibitem{Itoh et al.(1997)} N.~Itoh, A.~Nishikawa, S.~Nozawa et al.,
\emph{Contributions of the Plasmons to the Energy Density and Pressure in the Early Universe. II. Correlation Effects},
\emph{Astrophys. J.} {\bf 488} (1997) 507.



\bibitem[Tsallis(1988)]{1988JSP....52..479T} C.~Tsallis,
\emph{Possible generalization of Boltzmann-Gibbs statistics}, 
\emph{Journal of Statistical Physics} {\bf 52} (1988) 479.


\bibitem{Bertulani:2012sv} C.~A.~Bertulani, J.~Fuqua and M.~S.~Hussein,
\emph{Big Bang nucleosynthesis with a non-Maxwellian distribution},
\emph{Astrophys. J.} {\bf 767} (2013) 67 [arXiv:1205.4000].


\bibitem{Hou:2017uap} S.~Q.~Hou, J.~J.~He, A.~Parikh, D.~Kahl, C.~A.~Bertulani, T.~Kajino, G.~J.~Mathews and G.~Zhao,
\emph{Non-extensive Statistics to the Cosmological Lithium Problem},
\emph{Astrophys. J.} {\bf 834} (2017) no.2 165 [arXiv:1701.04149].


\bibitem{2019PhRvD..99d3505K} M.~Kusakabe, T.~Kajino, G.~J.~Mathews and Y.~Luo \ 2019, prd, 99, 043505.
\emph{On the relative velocity distribution for general statistics and an application to big-bang nucleosynthesis under Tsallis statistics}, 
\emph{Phys. Rev. D} {\bf 99} (2019) 043505 [arXiv:1806.01454].


\bibitem{McDermott:2018uqm}
S.~D.~McDermott and M.~S.~Turner,
\emph{Nuclear Kinetic Equilibrium During Big Bang Nucleosynthesis},
[arXiv:1811.04932 [hep-ph]].


\bibitem{Sasankan:2019oee}
N.~Sasankan, A.~Kedia, M.~Kusakabe and G.~J.~Mathews,
\emph{Analysis of the Multi-component Relativistic Boltzmann Equation for Electron Scattering in Big Bang Nucleosynthesis},
\emph{Phys. Rev. D} {\bf 101} (2020) no.12 123532 [arXiv:1911.07334].


\bibitem{1997PhRvL..79.2628O} M.~Opher and R.~Opher,
\emph{Was the electromagnetic spectrum a black body spectrum in the early universe?}, 
\emph{Phys. Rev. Lett.} {\bf 79} (1997) 2628 [arXiv:astro-ph/9708246].


\bibitem{2020PhPl...27b2106B} V.~B.~Bobrov, S.~A.~Trigger and I.~M.~Sokolov,
\emph{Spectral energy distribution of the equilibrium radiation and its asymptotic behavior in ideal gaseous plasmas}, 
\emph{Physics of Plasmas} {\bf 27} (2020) 022106.


\bibitem{2018arXiv181209472J} D.~Jang, Y.~Kwon, K.~Kwak and M.-K.~Cheoun, 
\emph{Big bang nucleosynthesis in a weakly non-ideal plasma},
\emph{Astron. Astrophys.} {\bf 650}, (2021) A121 [arXiv:1812.09472 [astro-ph.CO]].



\end{thebibliography}
\end{document}